\ifx\pdfvariable\undefined
\pdfoutput=1
\fi
\AtBeginDocument{\paperdetails{
  submitted=2017-04-01,
  published=2017-08-07,
  year=2018,
  volume=2,
  issue=1,
  articlenumber=2,
}}
\documentclass[english]{programming}
\usepackage[backend=biber]{biblatex}
\usepackage{textcomp}
\usepackage{hyperref}
\usepackage{graphicx}
\usepackage{multicol}
\usepackage{caption}
\usepackage{float}
\usepackage{proof}
\usepackage{multirow}
\usepackage{amsmath}
\usepackage{listings}
\usepackage{hyperref}
\usepackage{caption}
\usepackage[table]{xcolor}  %
\usepackage{color, colortbl}

\definecolor{pblue}{rgb}{0.13,0.13,1}
\definecolor{pgreen}{rgb}{0.0, 0.5, 0.0}
\definecolor{pred}{rgb}{0.9,0,0}
\definecolor{pgrey}{rgb}{0.46,0.45,0.48}
\definecolor{polive}{rgb}{0.23,0.39,0.30}
\definecolor{ppurple}{rgb}{128,0,128}
\definecolor{pbrown}{rgb}{0.59, 0.29, 0.0}
\definecolor{pcyan}{rgb}{0,255,255}

\begin{document}
\title{On the Effect of Semantically Enriched Context Models on Software Modularization}

\author[a]{Amir M. Saeidi}
\authorinfo[amir]{is a software engineer at Mendix and a PhD candidate at the Department of Information and Computing Sciences at Utrecht University. He is currently investigating various techniques to facilitate migration of legacy systems in financial domain to SOA. The techniques employed range from static analysis to data analysis to help both with understanding the legacy systems as well as their decomposition. He can be reached at \email{a.m.saeidi@uu.nl}.}
\affiliation[a]{Department of Information and Computing Sciences, Utrecht University, The Netherlands}
\author[a]{Jurriaan Hage}
\authorinfo[jurriaan]{is an assistant professor at the Department of Information and Computing Sciences at Utrecht University. His work in programming technology is largely focused on two aspects: the optimisation of functional languages by means of type and effect systems, and type error diagnosis for strongly typed functional languages. He is currently the lead maintainer of the Helium compiler for learning Haskell. Besides these two focus areas, he is also active in programming plagiarism detection, legacy system modernization, and the (soft type) analysis of dynamic languages. He can be reached at \email{j.hage@uu.nl}.}
\author[a]{Ravi Khadka}
\authorinfo[ravi]{is currently a technology architecture manager at Accenture. His focus area include legacy software modernization, light-weight architecture, model-driven development (MDD), API management,  and micro-service architecture. Khadka received his PhD in computer science from Utrecht University in $2016$. His PhD thesis is titled ``Revisiting Legacy Software System Modernization''. He can be reached at \email{ravi.khadka@gmail.com}.}
\author[a]{Slinger Jansen}
\authorinfo[slinger]{is an assistant professor at the Department of Information and Computing Sciences at Utrecht University. His research focuses on software product management and software ecosystems, with a strong entrepreneurial component. Jansen received his PhD in computer science from Utrecht University, based on the $2007$ work titled ``Customer Configuration Updating in a Software Supply Network'', PhD thesis Utrecht University. He can be reached at \email{slinger.jansen@uu.nl}.}
\paperdetails{
license=cc-by-nc,
  perspective=scienceempirical,
  area={Data mining and machine learning programming, and for programming, General-purpose programming},
}


\maketitle

\begin{abstract}
Many of the existing approaches for program comprehension rely on the linguistic information found in source code, such as identifier names and comments. Semantic clustering is one such technique for modularization of the system that relies on the informal semantics of the program, encoded in the vocabulary used in the source code. Treating the source code as a collection of tokens loses the semantic information embedded within the identifiers. We try to overcome this problem by introducing context models for source code identifiers to obtain a semantic kernel, which can be used for both deriving the topics that run through the system as well as their clustering. In the first model, we abstract an identifier to its type representation and build on this notion of context to construct a contextual vector representation of the source code. The second notion of context is defined based on the flow of data between identifiers to represent a module as a dependency graph where the nodes correspond to identifiers and the edges represent the data dependencies between identifiers. We have applied our approach to $10$ medium-sized Java projects, and show that by introducing contexts for identifiers, the quality of the modularization of the software systems is improved. Both of the context models give results that are superior to the plain vector representation of documents. In some cases, the authoritativeness of decompositions is improved by $66\%$. Furthermore, a more detailed evaluation of our approach on JEdit, an open source editor, demonstrates that inferred topics through performing topic analysis on the contextual representations are more meaningful compared to the plain representation of the documents. The proposed approach in introducing a context model paves the way for building tools that support developers in program comprehension tasks such as domain concept location and topic analysis.
\end{abstract}

\section{Introduction}
Traditional algorithms for lexical clustering of source code rely on the Vector Space Document (VSD) model, borrowed from the information retrieval field. In this model, the source code corpus is represented as a ``bag of words'' (BoW), without considering the relationships between the words. Using this representation, the similarity between any two modules is given as the inner product of the high-dimensional vectors indexed by the set of terms present in each module. This approach fails to include aspects of the formal semantics of the programming language, and hence is often inadequate to perform fine-grained source code analysis.

The bag of words model has well-known limitations: $1$) co-occurrence is the sole principle for inducing similarity between documents, and $2$) the same weight is given to different terms in the source code corpus. Although traditional preprocessing techniques for document vocabulary normalization such as eliminating stop words and stemming can be applied for preprocessing the source code corpus, its effect in this domain is still unclear.  On the other hand, the naming convention used in source code employs acronyms, domain-specific abbreviations and coded formats such as Hungarian coding~\cite{lawrie:2007:identifier, hill:2008:aam}. As a system evolves, the naming convention tends to become inconsistent throughout the source code~\cite{anquetil:1998:identifier}. 

To overcome the problems with the bag of words approach, many approaches have been proposed to incorporate the \emph{context} of a term in document representation. The n-gram model~\cite{brown:1992:ngram} is one such model that generalizes the BoW model by including all the ordered sequences of n words in the feature set, so it can capture relations between several adjacent words. In the n-gram model, the context is defined as a window of n words, surrounding the target word, whereas in the BoW model, the context of a word is the entirety of the corpus.  Even though the n-gram model considers the word order in short context, it suffers from data sparsity and high dimensionality. 

Developers may embed knowledge about application contextual information in identifier names. The notion of context of an identifier is different from that of the words appearing in natural language text, and approaches from information retrieval field cannot be directly applied to the source code.  Informally speaking, an identifier is a named program element, where it can be a namespace, class, method, variable or interface, that describes the role it may play in the execution of the program. Formally, an identifier is a placeholder for the set of values it may take at runtime. It is an abstraction of a memory location, where values are assigned to and read from. The data types and data structures of identifiers whether it is of a particular \emph{struct} in C, a \emph{copybook} in Cobol, or more generally a \emph{type} restricts the set of values it may take at runtime and is a good approximation of the context of an identifier. 

One model for incorporating context in the vector representation of the source program is to represent each module in terms of the pair of identifiers and its context, i.e. its type. In this representation, two modules are said to be similar if they have co-occurring identifiers that appear in the same context. Pairing each identifier with its type makes it possible to disambiguate polysemous identifiers based on the type they belong to. However, this representation may suffer from high-dimensionality, although in practice, the identifiers used in the source code don't appear in every context. Another representation involves abstracting an identifier in terms of its context, and represent a source code unit in terms of its types.

An alternative approach to the contextual vector representation is based on syntactic relations between the input word space. Syntax-based context models were proposed~\cite{grefenstette:1994:thesaurus, lin:1998a:similar} that go beyond mere co-occurrence of words to capture semantic relations based on syntactic relationships between words such as subject-verb. An identifier can be interpreted in combination with other identifiers that may have an effect on it. The flow of data in a program through assignment and value usages captures the data dependency relations between identifiers. The data dependency relations such as the definition-usage dependency relation, induces semantic relations between identifiers, irrespective of whether they are syntactically adjacent or not.  In this representation, a module is encoded in terms of a dependency graph capturing data dependencies derived from the syntactic structure of the model.

The equality of a context for induction of similarity between two identifiers is too strict, and can be relaxed by inferring semantic similarity between the contexts. Using the ontological structure of the type hierarchy, it is possible to enrich the context with semantic knowledge extracted from the semantic dependencies. Knowledge-rich methods are usually based on semantic networks~\cite{sridhara:2008:wordnet} or semantic-tagging of the corpus~\cite{falleri:2010:wordnet} to explicitly define the contextual meaning of a term in relation with other terms. Some researchers have adopted ontologies such as WordNet~\cite{hotho:2003:wordnetimproves} and Wikipedia~\cite{hu:2009:wikipedia} to enrich the representation of documents in the linguistic domain for cluster analysis.  As shown by Sridhara et al.~\cite{sridhara:2008:wordnet}, the general similarity measurements based on WordNet cannot be effectively applied in a software engineering context. Thus, we will automatically construct a knowledge base per software-system and measure the similarity in that domain. In this paper, we exploit semantic networks to represent the semantic knowledge of the system as a graph, capturing the similarity between the context and identifier names.

Our contributions in this paper are as follows:
\begin{itemize}
\item We propose two context models for identifiers in the source code, one based on a vector representation (CV) and the other using the data dependency graph (DG).
\item Orthogonally, we propose an approach to enrich the contexts and identifiers with semantic knowledge.
\item We make a comparison between the context models and establish the best semantically enriched context model for the clustering of software systems.
\item We observe the differences in the distribution of identifiers before and after introduction of contexts for source code identifiers.
\end{itemize}

We first survey related work in \autoref{sec:related} on various techniques for information extraction from source code and using that information for the clustering of software systems. In \autoref{sec:motivation}, we give a brief overview of existing context models used in the information retrieval field and outline our approaches to introduction of context models for identifiers in the source code. We proceed by defining the underlying context models in \autoref{sec:methodology}. In \autoref{sec:evaluation}, we show the validity of our approach by performing clustering on semantically enriched contextual representation of the source program and make a comparison between the two models using $10$ medium-sized real-world Java programs. We conclude in \autoref{sec:conclusion} and outline future work.

\section{Related Work}
\label{sec:related}

We classify related work into three categories, based on the research problems we investigate in this paper. In the first category, we review related work that utilise knowledge-rich methods to infer semantic similarity between word-pairs in the source code. In the second category, we survey approaches that rely on extracting lexical terms based on the context they appear in the source code to perform program comprehension tasks. Finally, we outline approaches that rely on the vocabulary found in the source code to modularize a software system. 

\subsection{Inferring similarity between lexical terms}

Falleri et al.~\cite{falleri:2010:wordnet} propose an approach to automatically construct an ontology from the software system, by extracting concepts from identifier names in the source code and organizing the identifiers into a WordNet-like structure. Their approach to building such structure comprises of techniques such as tokenization of names and part-of-speech tagging. Tian et al.~\cite{tian:2014:database} propose a technique that leverages information from software information sites to construct a similarity relation between software terms. The semantic relations for the software term set is estimated based on the co-occurrence information extracted from StackOverflow. They demonstrate that this technique is significantly more effective than a WordNet-based similarity measure. Abebe and Tonella~\cite{abebe:2015:domainconcepts} exploit structural and linguistic information in the source code to extract domain facts and store them in an ontology. The extracted ontology can be used by developers to support them in program understanding tasks such as domain concept location. Similar to these techniques, we automatically construct a taxonomy of words extracted from the public API, computed in a software system-specific fashion. 

Mahmoud and Bradshaw~\cite{mahmoud:2015:relatedness} investigate the performance of several semantic relatedness measurements from the natural language processing field. They observe that corpus-based methods outperform methods that rely on external semantic knowledge. To further improve the relatedness measure, they propose an information-theoretic method called \emph{Normalized Software Distance} that exploits the distributional cues of identifiers across the system. Similar to this approach, we employ semantic relatedness measurements to compute semantic similarity between lexical terms to enrich identifiers extracted from the corpus of the source code. In addition, graph kernels are used to calculate semantic similarity between identifiers in the constructed ontology.

\subsection{Contextual representation of words in source code}

Hill et al.~\cite{hill:2011:phrasal} leverage phrasal concepts (PCs) to improve software search. They introduce a relevance scoring function that integrates information about the position of query words in the code as well as the semantic role of query words in phrasal concepts. They show that location of a query term in the method signatures and method bodies improves the accuracy of the search results. Another technique that leverages the context of words in comments and code is proposed by Yang and Tan~\cite{yang:2012:context, yang:2014:swordnet} to automatically discover semantically related words in software systems. Their hypothesis is that if two words or phrases are used in the same context in comment sentences or identifier names, then they may exhibit syntactic and semantic similarity. The semantically related words are then used for code searching. The effectiveness of their approach is demonstrated using seven Java and C code bases. Howard et al.~\cite{howard:2013:mine} take a similar approach by extracting action verb pairs from method names and their associated comments. The assumption they make is that action verbs present in the method signature and the header comment are semantically related. These two approaches heavily depend on the quality of naming and the descriptions provided in the header comments of methods. These approaches are limited to a notion of context of words from the natural language perspective, and does not concern the context of the identifier name as a whole within a program.

Another model for suggesting method and class names is proposed by Allamanis et al.~\cite{allamanis:2015:names} that goes beyond the local context of the named element to suggest functionally descriptive names. They introduce neural probabilistic language model for source code that learns which names are semantically similar by embedding them into a high-dimensional space. They show that names with similar embeddings tend to be used in similar contexts. In contrast to this work where semantic information about tokens are learned only from statistical co-occurrences of tokens, we define explicit context models that capture semantic similarity between the tokens within the source code.

Zilberstein and Yahav~\cite{zilberstein:2016:semantic} propose an approach for measuring semantic similarity between code fragments by performing static analysis on code snippets to extract their type signatures and incorporating their associated natural language descriptions. We adopt a similar approach by performing syntactic analysis to compute an abstraction of an identifier in terms of its type, and augment it with its natural language description to compute the similarity between source code documents.

\subsection{Semantic clustering of software systems}

Semantic clustering, proposed by Kuhn et al.~\cite{kuhn:2007:semantic} is a modularization technique based on Latent Semantic Indexing that partitions the software system based on a common use of vocabulary, and identifies linguistic topics within the source code. The observation they make is that the identified topics mainly correspond to application concepts and architectural components, but fails to reveal the domain semantics. Corazza et al.~\cite{corazza:2010:probabilistic,corazza:2011:lexical} introduce a weighing mechanism to distinguish between the information extracted from different zones in the source code, such as comments and class names. Santos et al.~\cite{santos:2014:remodularization} investigates conceptual metrics and semantic clustering in the context of six real-world modularization projects. They conclude that latent topics within the source code become more coherent after restructuring of the software system. We have adopted semantic clustering to perform modularization on semantically enriched contextual representation of software system and evaluate our approach on $10$ Java open source projects.

\section{Motivation}
\label{sec:motivation}
The basic premise in contextual approaches in natural language processing (NLP) is that two words are closely associated if they share the same context. For instance, two words that are associated with objects that `fly' and `lay eggs' are more closely related than the ones associated with `swimming', even though we don't know what exactly they are. Similar to word similarity, two documents are said to be similar if they contain the same words in context. The N-gram model is one representation that tries to incorporate a window of words in the meaning of a target word. %

The definition of linguistic context is different from the context of identifiers in software systems. Identifiers convey information about the values they may hold during the execution of the program. Identifiers are user-defined referenceable elements that are uniquely identified by the compiler, which can take one of the following forms: $1$) class/module names, $2$) fields/properties, $3$) method/procedure/function names and their $4$) parameters, and $5$) local variables. The use of syntactic analysis paves the way for introducing more sophisticated context models for identifiers than approaches based on word co-occurrence or co-occurrence within a window of words. Syntactic analysis of source code allows us to extract contextual information about identifiers such as dependency relationships between identifiers and their type information.

In the following, we outline two models for incorporating the contexts of identifiers in the representation of the source programs. We proceed by describing how the identifiers and the contexts can be enriched with semantic information, extracted from implicit semantic dependencies as well as the similarity between their lexical representations.

\subsection{Contextual Vector Representation}
Traditional models for document similarity are based on a document-term matrix. In this model, a document vector represents the collection of words occurring in a document (also known as bag of words). In a bag of words representation of the document corpus, the mere co-occurrence of words is used to induce the document similarity. The implicit context of word in this approach is the entirety of the documents corpus\footnote{Although there is a global context, we will refer to these approaches as context-free.}. This context free representation of the source program amounts to the bag of identifiers (BoI), comprising of the collection of identifiers present in the source code unit.

The majority of works in NLP on incorporating context in the representation of document concerns the use of the local context of a word for determining its meaning. Local context is usually represented as a window of words around the target word, regardless of the distance or grammar relations between the words. This representation can't be directly applied to programming languages. In contrast to natural languages where adjacent terms may exhibit semantic relations, in programming languages, how identifiers are grouped together in the concrete syntax of the program is usually different from how they are semantically related. To clarify further, let us look at the following simple example:
\begin{lstlisting}
int a; boolean b;
\end{lstlisting}
From the declaration statements above, no semantic relations can be inferred between identifiers $a$ and $b$, even though they appear adjacent to each other. One context model that captures the possible semantic relations between identifiers is to incorporate the data type of an identifier into the vector representation of the source code unit.

\subsection{Syntax-based Context Models}
An alternative to the vector-based representation of contexts are the syntax-based models. In these syntactic models, contexts are defined over words that have a syntactic relationship with the target words of interest.  This makes semantic spaces more flexible: different types of contexts can be selected; words do not have to co-occur within a small, fixed word window; and word order or differences in text structure can be naturally reflected in the semantic space~\cite{pado:2007:dependency}. The hypothesis is that syntactic structure is a close reflection of lexical meaning. In general, syntactic models capture a stricter notion of context, and should capture tighter semantic relations between words.

In programming languages, the concrete syntax of a program is used by a developer to specify a desired behaviour through an interaction between program elements. The syntactic relations between these elements are extracted during the parsing stage of compilation to build an abstract structure of the program (AST).  This structure is then mapped to a representation that can be ``interpreted'' by a runtime engine.  Hence, the abstract syntax serves as a bridge between the concrete syntax of a program and its meaning.  Any approach to find relations between identifiers cannot merely depend on textual or syntactic structure of the program, but has to consider semantics too. Our approach is based on program dependence graphs which builds on the abstract syntax to represent the structure of a program and the data flow within it. 

Program dependencies are syntactic relationships between program statements, which are used to obtain an approximation about the flow of information between program elements. There are two types of program dependencies: control dependencies, which are features of a program's control structure, and data flow dependencies, which denote variable definitions and usages in a program. Here in this paper, we restrict ourselves to flow-insensitive data dependencies, and don't take into account control dependencies between different program statements.
 
To further clarify the data dependencies in a program, let us consider the following example:
\begin{lstlisting}
a = b + c;
\end{lstlisting}

The memory location identified by the name $a$ is over-written by a value that depends on the values of the memory locations identified by $b$ and $c$. Hence, there is a data dependency from identifiers $b$ and $c$ to identifier $a$.

\subsection{Enrichment of Context Models with Semantic Knowledge}
In the aforementioned context models, the contexts or identifiers must be equal to induce any similarity between the documents that contain them. This restriction can be relaxed by inferring a semantic similarity between pairs of identifiers or contexts. For instance, the identifiers `developer' and `secretary' are of two different types, but both of their types share the same superclass called `Employee'. If external knowledge such as word relatedness is available, it is possible to enrich the document representation using this semantic information. This information can be provided through supervision by an expert in the domain, or derived from a knowledge base. Hu et al.~\cite{hu:2009:wikipedia} leverage Wikipedia concept and category information to embed semantic information into the text word vector. %

The most obvious approximation of the context of an identifier is its type, or the data structure it belongs. A data type or a data structure constrains the set of values an identifier may take during runtime. For instance, consider the identifiers: `HashSet', `OrderedSet' and `setName'. Although the term `set' is shared between all the three,  the last one is unrelated to the first two. The first two are subtypes of the `Set' class, whereas the last one is a setter method. On the other hand, `ArrayList' and `HashSet' are lexically unrelated, but both are subclasses of the Collection class.

The measure of semantic similarity between contexts and identifiers should be sensitive to:
\begin{itemize}
\item Conceptual similarity, measuring similarity between the concepts, i.e. types, and using it to measure the closeness between identifier names based on whether they belong to the same concept/structure.
\item Lexical similarity, measuring similarity of identifier names based on their lexical representation.
\end{itemize}

The idea behind conceptual similarity is based on the observation that two identifiers are more similar if they have a \emph{similar} context, that is they belong to a similar type. Before we can compute the similarity between two identifiers, we need to establish what it takes for two types to be similar. In our first approach, we will employ semantic relatedness measurement metrics from the information retrieval field that computes similarity between senses of words.  The second approach computes the similarity between senses of words by propagating the information in the semantic network. 

Conceptual similarity is not the sole determinant in inducing semantic similarity between identifiers. Two identifiers could be of compatible types, but have no semantic relationships. For instance, identifiers  `PersonName' and `CarModel' are of type String, but have different senses. Another approximation for the context information is the lexical similarity between two identifiers. For more information about computing the lexical similarity between identifier names, please refer to \autoref{apx:sec:lexical}.

 \section{Methodology}
\label{sec:methodology}

In this section, we describe our approach to building various context models for the representation of the vocabulary found in the corpus of source code. Before outlining the context models, we briefly describe the process for enrichment of identifiers and contexts with semantic knowledge.

\subsection{Semantic Enrichment Process}

\begin{figure*}
\begin{center}
\includegraphics[width=11cm]{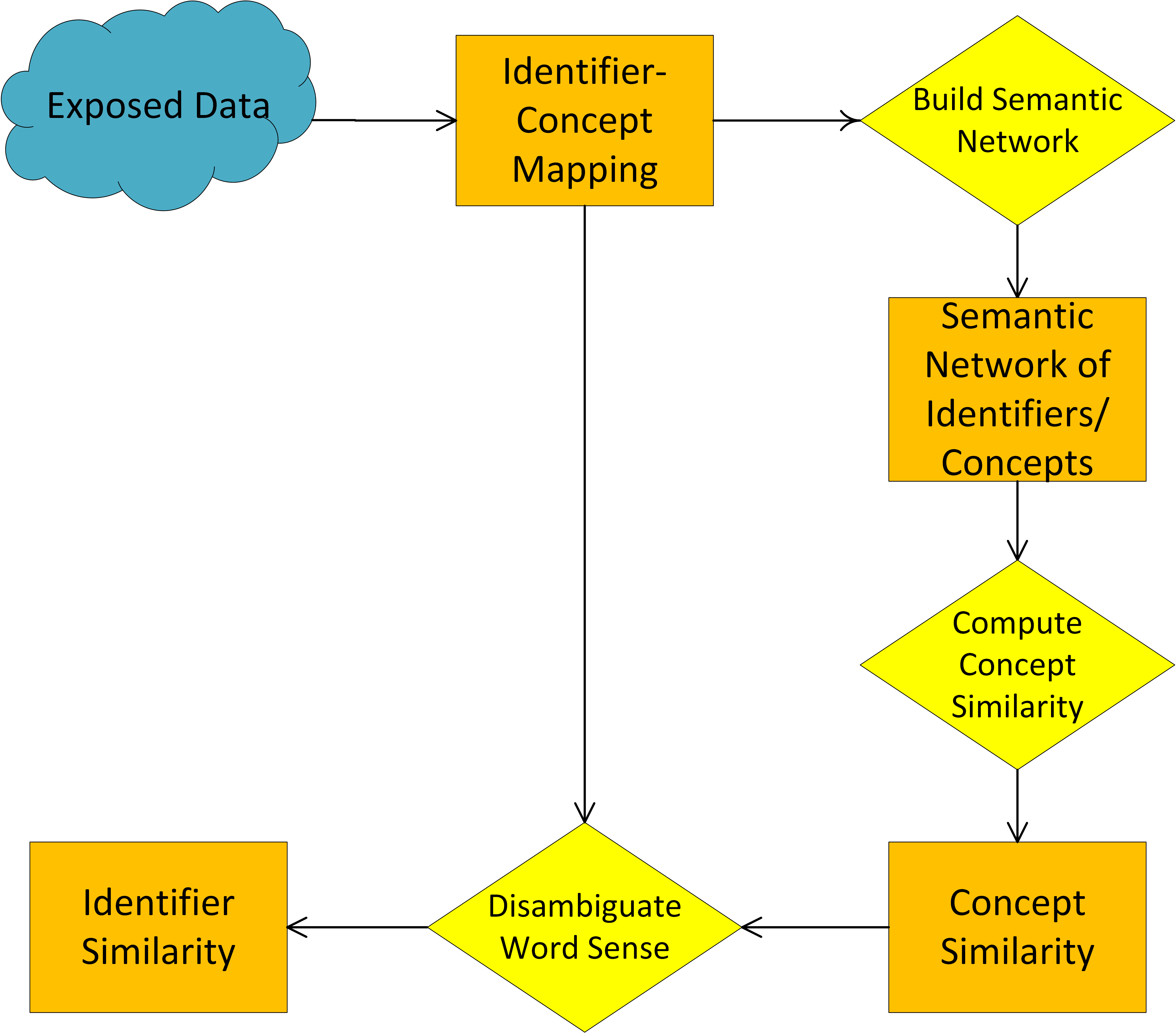}
\caption{Semantic Enrichment Process}
\label{fig:enrichment:process}
\end{center}
 \end{figure*}

The semantic enrichment process is depicted in \autoref{fig:enrichment:process}. In the first step, we construct a knowledge base (KB) for representing the semantic dependencies of identifiers and their associated contexts (identifier-concept mappings). We scan through the source code to identify semantic dependencies, i.e. the packages imported in each class.  Using the established imported libraries, we automatically extract referenceable identifiers exposed in a public API to construct the semantic network. This extraction is performed through shallow parsing of the source code to only retrieve the relevant public members both in internal and external sources. Since we focus on unsupervised extraction of information from the semantic dependencies, we have no notion of synonymy and antonymy. However, this information can be provided through supervision by a developer or an architect who is familiar with the source code naming.  For more information about how to automatically construct a semantic network from the software system, please refer to \autoref{apx:sec:conceptual}.

Once the knowledge base is constructed, we employ two techniques to compute the similarity between concepts, one based on employing semantic relatedness measures from NLP, and the other using the diffusion kernel to compute the similarity between nodes in the semantic network. Once the concept similarity is computed, it is possible to use the identifier-concept mappings to disambiguate the meaning of identifiers. Each identifier may have different meanings (or senses), depending on the type it may take. For instance, the identifier `tree' may represent a plant or a data structure, or something unrelated to its textual meaning, based on the type it belongs to. Like word sense disambiguation in NLP, we perform identifier sense disambiguation to measure the similarity scores between identifiers. In \autoref{apx:sec:conceptual}, we thoroughly describe the aforementioned techniques for enriching different context models with conceptual similarity.

Depending on how the context is modelled, various representations can then be enriched with semantic information, which can be used to perform cluster analysis for tasks such as software modularization. The semantic similarity measures only account for the exposed members of the API. Hence, any program element referenced within the body of the method which is scoped locally to that method cannot be annotated with the semantic information. 

\subsection{Contextual Vector Representation}
As stated earlier, the plain BoW representation of the documents is a vector space model where the entirety of the corpus is the context for each word. The bag of features (BoF) is the generalization of the BoW, representing the frequency of features per document unit. The bag-of-features model is defined as follows:
$$
\phi(d) = c(tf(f_{1}), tf(f_{2}), \ldots, tf(f_{n}))
$$
where $tf(f_i)$ denotes the frequency of feature $f_i$ in document $d$. The features can be the identifiers, identifier-context pairs, or just contexts. When the features are plain words, the BoF is equivalent to the BoW representation.

A problem with the BoW representation is the existence of polysemy in identifier names, leading to imprecision in the similarity relations between documents. We propose two variants of the vector-based context model that take the type of identifiers into account. The first variant is to augment each identifier with the context it appears within the source code, giving rise to the context-identifier vector representation of the corpus. The possible space for this representation is the cartesian product of the identifiers and contexts (i.e. types). This contextual representation is a sparse matrix, since in practice, not every identifier takes every possible type. Another contextual vector representation is the abstraction of identifiers in terms of their context, i.e. the data types. In this representation, each source code unit is mapped into a space where each feature is a data type and the feature's value is the frequency of occurrence of that type in the source code unit. 

To further clarify the various contextual vector representation of the source code, let us consider the following example.

\begin{lstlisting}[caption=An example in Java, frame=tlrb, label=lst:employee:snippet, basicstyle=\fontsize{8}{9}\selectfont\ttfamily]{Name}
public class Employee {
    public String name;
    public double salary; 
    public Date hireDay;
         
    public Employee(String name, double salary, int year, int month, int day) { 
    	this.name = name;
    	this.salary = salary;
    	Date temp = new Date(year, month, day);
    	this.hireDay = temp;
    }  

    public double raiseSalary(double byPercent, double bonus) {
      double temp = salary * byPercent / 100;
      salary += temp + bonus;
   }
}
\end{lstlisting}

Table~\ref{tbl:context:vector:representations} gives various contextual vector representations of the source code corpus used in this paper, starting from the context-free representation (i.e. bag of identifiers), to bag of identifier-contexts, as well as bag of contexts representation. The identifier `temp' as used in the code snippet above, takes two forms, one as an instance of `Date' in the constructor, and the other as a `double' in the body of the method `raiseSalary'. The identifier-context representation of the class, as shown in \autoref{tbl:context:vector:representations} can distinguish between these two identifiers by incorporating contextual information, the information which is lost in the plain bag of identifiers representation.

\begin{table} 
\centering 
\caption{From left to right: $1$) Bag of identifiers in class `Employee', $2$) Bag of pairs of identifer-types in class `Employee', and $3$) Bag of types in class `Employee'}
\begin{tabular}{ccc}
\begin{tabular}[t]{c c}
&Employee\\\hline
name&$2$\\
salary&$4$\\
hireDay&$1$\\
year&$1$\\
month&$1$\\
day&$1$\\
temp&$4$\\
byPercent&$1$\\
bonus&$1$\\
\end{tabular}
&
\begin{tabular}[t]{c c}
&Employee\\\hline
(name,{\color{blue}String})&$2$\\
(salary,{\color{blue}double})&$4$\\
(hireDay,{\color{blue}Date})&$1$\\
(year,{\color{blue}int})&$1$\\
(month,{\color{blue}int})&$1$\\
(day,{\color{blue}int})&$1$\\
(temp,{\color{blue}Date})&$2$\\
(byPercent,{\color{blue}double})&$1$\\
(bonus,{\color{blue}double})&$1$\\
(temp,{\color{blue}double})&$2$\\
\end{tabular}
&
\begin{tabular}[t]{c c}
&Employee\\\hline
{\color{blue}String}&$2$\\
{\color{blue}double}&$8$\\
{\color{blue}int}&$3$\\
{\color{blue}Date}&$3$\\
\end{tabular}
\end{tabular}
\label{tbl:context:vector:representations}
\end{table}

In \autoref{apx:sec:semantickernel}, we demonstrate how to enrich the vector representation $\phi(d)$ with semantic knowledge to build a semantic kernel. The semantic kernel leaves composite identifier names unbroken, it captures the semantic closeness of synonymous identifier names, and performs word sense disambiguation for polysemous identifiers.

\subsection{Dependency-based Construction of Context Model}
The dependency-based context model is based on computing the similarity between two modules based on how much their corresponding data dependency graph is similar. It employs the data dependency graph to represent the semantic dependencies, and therefore considers the data flow between program identifiers (as an abstraction of the semantics). The data dependency graph makes an explicit representation of the definition-use relationships implicitly present in a source program. %

The grammar shown in \autoref{fig:kcprogram} describes a simple language, which we use here to demonstrate how a data dependency graph can be constructed from different constructs present in the program.

\def \Exp{\mathbf{e}}
\def \module{\mathbf{M}}
\def \statement{\mathbf{S}}
\def \identifier{\mathbf{Id}}

\newcommand{\funci}[2]{\mathit{fun_{#1}(} \mathit{#2} \mathbf{)} }
\newcommand{\func}[1]{\mathbf{fun(} \mathit{#1} \mathbf{)} }
\newcommand{\ret}[1]{\mathbf{ret} \mbox{ } \mathit{#1}}
\newcommand{\call}[2]{\mathbf{call} \mbox{ } \mathit{#1} \mathbf{(} \mathit{#2} \mathbf{)}}

\begin{figure}
	\centering
	\[
	\begin{array}{lrcl} 
\mbox{module} &  \module & ::= & \funci{i}{\overrightarrow{v}}\\
\mbox{functions} & \func{\overrightarrow{v}} & ::= & \overrightarrow{s}\\
\mbox{statements} & \statement & ::= & e\\
&&|& v := e\\
&&|& \ret{e}\\
\mbox{expressions} &  \Exp & ::= & id \\
&&|& e.id\\
\mbox{idenitifiers} & \identifier & ::= & v\\
&&|& \call{f}{e}\\
\end{array}
\]
\caption{The context-free grammar used to construct the data dependency graph}
\label{fig:kcprogram}
\end{figure}

Suppose $\module$ is the set of modules with a vocabulary set $D = \{ id_1,id_2, \dots, id_n\}$, where $id_i$ represents the $i^{th}$ unique identifier present in $\module$. The data dependency graph $G$ for the module $m$ is denoted as a directed graph $G = <V,E>$. Similar to a PDG and the AST, each identifier $id_i$ in the vocabulary set is mapped to a vertex in the graph. $E$ is the collection of edges between the vertices in the graph, where each edge $e=(v_i, v_j)$ indicates that there is a data dependency from $v_i$ to $v_j$, that is the value stored in $v_j$ may affect the value computed from $v_i$, at some point in the program. The meaning of an identifier $id$ is sought in the context of 1) the identifiers which their values reach the identifier $id$, or 2) the identifiers that may be affected by the value of the identifier $id$.

Functions are usually called using prefix notation where the function name is followed by its arguments. However, for some function calls such as arithmetic and logical operators, they appear in infix notation. We transform these operators into prefix notation, where the operator is replaced with a nameless function call\footnote{We are only interested in the flow of data between variables, and the type of the operation for in-built operators can be erased.}. Each variable and function call is attributed to a vertex in the data dependency graph. 

Data definition edge represents the assignments of values to variables. The definition edges are similar to the value dependence edges, except that they show that a computed value is stored into a variable. The return statement assigns the value stored/computed in an identifier to the container function. In function calls, the arguments (actual parameters) are passed into the function, so there are edges from the arguments to the function (formal) parameters. We assume that the arguments are passed by value, hence the flow is uni-directional. Between the components of an expression in an access path, there are bidirectional edges between target variables. Consider the example `employee.setVehicle(newCar);' . The receiver object may be affected (a conservative approximation) by the side-effects arising from calling the method `setVehicle'. On the other hand, the `setVehicle' may use values stored in the receiver object `employee'. In our analysis, we do not distinguish between instance and class members. Another shortcoming of our analysis is that we do not take into account name aliases, although the aliasing operation maps to a single data dependency edge.

\begin{figure}
\begin{center}
\includegraphics[width=12cm]{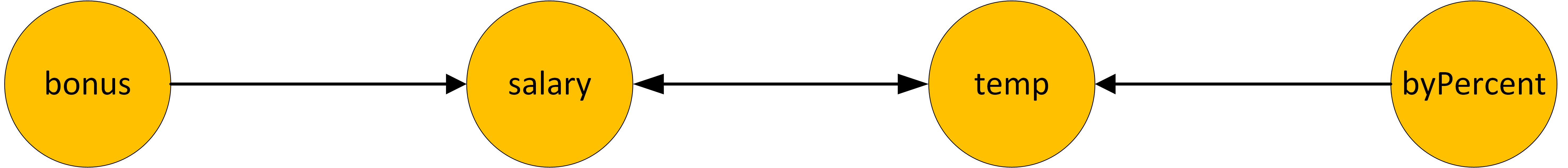}
\caption{The data dependency graph of the Java example program, composed of abstract nodes designating the identifier names and abstract links designating different types of data flow.}
\label{fig:salary:graph}
\end{center}
 \end{figure}

For example, let us have another look at the code snippet in \autoref{lst:employee:snippet}. The `raiseSalary' method in Class `Employee' calculates the new salary based on the amount of `raise' and end of year `bonus', and assigns the new calculated amount to the field `salary'. Figure~\ref{fig:salary:graph} depicts the data dependency graph constructed for the above code snippet. 

Graph kernels~\cite{vishwanathan:2010:graph} provide a natural way to compute the similarity between graphs. The random walk graph kernel is one such technique that has been used for classification and measuring similarities between graphs. Given two graphs, the random walk kernel computes the number of common walks between two graphs. Two walks are common between the graphs if the lengths of the walks are equal, and the label sequences are the same. For more information about graph kernels, please refer to \autoref{appendix:sec:graphkernels}.

The random walk kernel is parameterized over a positive semi-definite function that computes the similarity between the labels in the graph, i.e. the identifier names. We can restrict the label similarity to return $1$ when the labels are identical, and $0$ otherwise. However, this notion of similarity is too strict and doesn't capture the similarity induced from conceptual and lexical similarity between the identifier names. For instance, the two identifier names, `salary' and `raiseSalary' are not identical, but are conceptually and lexically similar. We use the semantic kernel produced from the semantic enrichment process to adjust the label similarity based on both the conceptual and lexical similarity of identifier names, hence making this context model semantics-aware.

\section{Empirical Evaluation}
\label{sec:evaluation}
In this section, we present the empirical evaluation to assess the effectiveness of our proposed context models for semantic clustering of software systems. We have conducted experiments on $10$ medium-sized open source Java projects~\cite{saeidi:2017:dataset}, as given in \autoref{tbl:benchmark}, using different encodings of the semantically enriched context models and the results show that our approach is more effective in semantic clustering than previous approaches.

\begin{table*}[t]
\begin{center}
 \caption{The benchmark for empirical evaluation of lexical clustering of various semantic space models} 
    \begin{tabular} {l l r r r}
    \rowcolor{gray!50}
    \hline
     \textbf{System} & \textbf{Description} & \textbf{Version} & \textbf{\#Classes}  & \textbf{KLOC} \\
		 Apache Ant & building library & $1.9.3$ & $691$ & $31.07$ \\ 
		 Apache Hadoop & distributed computing library & $0.20.2$ &$707$ &$102.53$ \\ 
		 Apache Log4j & logging library & $1.2.17$ & $218$ &$10.54$  \\ 
		 Eclipse JDT Core & Java Development Tools& $3.8$ & $1,276$  & $162.59$ \\ 
		 JDOM & XML library & $2.0.5$ & $140$ & $21.93$ \\ 
		 JEdit & text editor & $5.1.0$ & $536$ & $112.58$  \\ 
		 JFreeChart & chart library & $1.2.0$ & $655$ & $100.05$ \\ 
		 JHotDraw & GUI framework & $7.0.6$ & $284$ & $12.01$ \\ 
		 JUnit & unit testing &$4.12$ & $167$ & $6.09$ \\ 
		 Weka & machine learning library  & $3.6.11$ & $1,346$ & $203.51$\\ \hline
	\end{tabular}
\label{tbl:benchmark}
\end{center}
\end{table*}

\subsection{Methods and Evaluation Measures}
Lexical-based clustering (semantic clustering) is employed as a basis to evaluate the performance of the methods to enrich the source code terms. Software clustering is the process of decomposing software systems into more cohesive and maintainable subsystems. Semantic clustering groups source code modules into clusters based on common usage of vocabulary.

The quality of the produced clusters is evaluated against an authoritative decomposition. We have adopted the approach employed in works~\cite{corazza:2010:probabilistic, corazza:2011:lexical, wu:2005:comparison} to use the package structure as the authoritative decomposition of the system. In contrast to some of the aforementioned approaches, we have decided not to flatten the package structure and preserve its taxonomy. In a package structure, each package is a well-defined unit which encapsulates its subpackages. Hence, flattening the package structure loses the scope of packages, and is not a good ground truth for evaluating the clustering. For instance, a `util' subpackage in `model' subdirectory is more closely associated with its sibling subpackages than subpackages in `GUI' subdirectory. Furthermore, to ensure integrity across the clusters, we eliminate packages that have 4 or fewer modules and break the packages with more than $40$ classes into smaller packages. 

Hierarchical clustering is a clustering technique that enables us to produce a hierarchy of clusters. The advantage of using hierarchical clustering for modularization of system modules is that:
\begin{itemize}
\item It does not assume a particular value for the number of clusters, as needed by partitioning algorithms such as k-means.
\item The generated tree closely resembles the package structure of the software system. The tree can be cut into several groups by specifying the desired number of clusters.
\item Hierarchical clustering algorithms (linkage algorithms), unlike partitioning algorithms such as k-means, are deterministic, and produce the same decomposition every time. 
\end{itemize}

Since hierarchical clustering algorithms work on distance matrices, we compute the corresponding distance matrix from the similarity matrix by subtracting it from $1$, i.e. $D = 1- S$.

To evaluate the (dis)similarity between the produced hierarchical clustering and the authoritative hierarchical decomposition, we have opted for the following scores. Any (dis)similarity score between the trees should be sensitive to the taxonomic structure of the trees.

\subsubsection{Tree Edit Distance}
Tree edit distance (TED) measures the minimal number of edit operations required to transform one tree to another. The edit operations on labelled trees include renaming of a node, removal of a node and connecting its children to its parent, and addition of a new node. We have implemented the tree edit distance algorithm described in~\cite{zhang:1989:ted} and apply it to the trees corresponding to the produced clustering and the package structure. The only constraint we impose in our implementation of the metric is that renaming of leaf nodes (i.e. modules) is not allowed, while renames of non-tip nodes (i.e. subdirectories) incur no costs. All ther other operations have a single unit cost.

\subsubsection{Path Difference Metric}
The path difference metric (PD) is a distance metric proposed by Steel and Penny~\cite{steel:metrics:1993} that measures the topological dissimilarity of two trees, while ignoring the lengths of branches. In this metric, for each possible pair of leaves in each tree, the length of shortest paths (the number of nodes traversed from one node to the other) are computed.  The sum of the differences in these shortest path lengths is the distance between the two trees. In the variant proposed by Steel and Penny~\cite{steel:metrics:1993}, the terms in the difference are raised to the power of two. The lower value for PD indicates a higher quality decomposition.

\subsubsection{Complete Linkage Algorithm}
We have conducted a set of experiments using different hierarchical clustering algorithms to find a linkage algorithm which gives the most authoritative solutions. The trials show that the complete linkage algorithm yields results that are superior to other linkage algorithms. The complete link aggressively forces clusters to be as dissimilar as possible. The result of complete link is also consistent with our objective of decomposing the system to subsystems that are cohesive without negatively affecting coupling.
  
\subsection{Experimental Settings and Results}
 
In this evaluation\footnote{The implementation of the approaches used in this paper can be found as part of the \emph{GeLaToLab} project at: \url{https://github.com/amirms/GeLaToLab}}, we first establish the best choice for contextual vector representation of the source code, by making a comparison between the aforementioned models. Once the best model is identified, we evaluate our two semantically enriched context models for the source code representation, by comparing it against the plain bag-of-identifier representation. 

In \autoref{apx:sec:evaluationenrichment}, we have conducted the set of experiments to find the best technique for semantic enrichment of context and identifiers using conceptual and lexical similarity. The results demonstrate that the normalized longest common substring (LCS) is a good choice for computing the similarity between identifier names. LCS measures the length of the longest consecutive set of characters shared between the two identifier names. On the other hand, diffusion of similarity over the semantic network (SN) yields results that are superior to semantic relatedness measures for computing similarity between contexts and identifiers. In contrast to semantic relatedness measures, the diffusion kernel doesn't distinguish between different kinds of relationships in the SN, and induces less similarity through nodes with high (in/out)degree such as common utility and helper functions.

\subsubsection{Evaluation of Context Vector Representation}
We first make a quantitative comparison between different contextual vector representations to find the best method in this context model. 

We compare the performance of the following three methods:                                                                                                                                                                            
\begin{itemize}
\item EBoI: The bag of identifiers enriched with the semantic matrix from element-wise product of diffusion-based conceptual similarity and normalized LCS kernel with the idf weighing schema.
\item EBoIT: The bag of identifier-types enriched with the semantic matrix from the Kronecker product\footnote{Given two real matrices $A \in R^{n\times m}$ and $B \in R^{p \times q}$, the Kronecker product $A \otimes B \in R^{np \times mq}$ is a block matrix defined as the product of each element in A with all the elements in B.} of diffusion-based conceptual similarity and normalized common substring kernel with the idf weighing schema.
\item EBoT: The bag of types enriched with the semantic matrix from diffusion-based conceptual similarity over type structure with the idf weighing schema.
\end{itemize}
               
\begin{table}
\begin{center}
          \caption{The comparison of plain bag of identifiers with enriched variants of bag of identifiers, bag of identifier-types, and bag of types} 
          \begin{tabular}{  c  c  c  c  c  c  c  c }
    \hline
    \multirow{2}{*}{\textbf{System}} &  \multicolumn{2}{c}{\textbf{EBoI}}  &  \multicolumn{2}{c}{\textbf{EBoIT}} &  \multicolumn{2}{c}{\textbf{EBoT}} \\\cline{2-7}
    & \textbf{PD}  &  \textbf{TED}  & \textbf{PD}  &  \textbf{TED}  & \textbf{PD}  &  \textbf{TED}  \\ \hline
Apache Ant&\cellcolor[gray]{0.8}{$7457$}&\cellcolor[gray]{0.8}{$1961$}&$8738$&$2049$&$13006$&$2049$\\ \hline 
 Apache Hadoop&$2466$&$711$&\cellcolor[gray]{0.8}{$2369$}&\cellcolor[gray]{0.8}{$682$}&$2887$&$733$\\ \hline 
 Apache Log4j&$1365$&$457$&\cellcolor[gray]{0.8}{$1112$}&\cellcolor[gray]{0.8}{$432$}&$1731$&$483$\\ \hline 
 Eclipse JDT Core&$14344$&$2374$&\cellcolor[gray]{0.8}{$11057$}&\cellcolor[gray]{0.8}{$2235$}&$18382$&$2628$\\ \hline 
 JDOM&$765$&$265$&\cellcolor[gray]{0.8}{$640$}&\cellcolor[gray]{0.8}{$247$}&$784$&$265$\\ \hline 
 JEdit&\cellcolor[gray]{0.8}{$4461$}&\cellcolor[gray]{0.8}{$1194$}&$5305$&$1232$&$5814$&$1224$\\ \hline 
 JFreeChart&$5652$&$1380$&\cellcolor[gray]{0.8}{$5618$}&\cellcolor[gray]{0.8}{$1352$}&$6531$&$1450$\\ \hline 
 JHotDraw&$1958$&$653$&\cellcolor[gray]{0.8}{$1946$}&\cellcolor[gray]{0.8}{$651$}&$2304$&$661$\\ \hline 
 JUnit&\cellcolor[gray]{0.8}{$563$}&\cellcolor[gray]{0.8}{$252$}&$697$&$288$&$784$&$296$\\ \hline 
 Weka&$8471$&$2038$&\cellcolor[gray]{0.8}{$7362$}&\cellcolor[gray]{0.8}{$1992$}&$9925$&$2165$\\ \hline  
	\end{tabular}
\label{tbl:comparative:cv}
\end{center}
\end{table}     

Table~\ref{tbl:comparative:cv} gives the results of our contextual vector model against the enriched BoI representation of the source code corpus. In $7$ out of $10$ cases, the bag of identifier-types (BoIT) give results that are superior to the other representations, however, the bag of types (BoT) scores lower than even the enriched BoI. This demonstrates the significance of the information that identifier names contain, that should be exploited in any lexical-based program analysis. This finding shows that the full context of an identifier is best expressed in terms of the pair of identifier name and its type.
 
\subsubsection{Comparison of Context Models}
Based on the previous results, we have configured our enrichment approaches for both the string kernel and the choice of conceptual similarity measurement. We compare the performance of the following three methods:                                                                                                                                                                            
\begin{itemize}
\item BoI (baseline): The plain bag of identifiers (BoF) with the idf weighing schema.
\item $SSK_1$: The bag of features enriched with the semantic matrix from element-wise product of diffusion-based conceptual similarity and normalized common substring kernel with the idf weighing schema.
\item $SSK_2$: The semantic kernel computed from performing random walk kernel between the data dependency graphs of modules with element-wise product of diffusion-based conceptual similarity and normalized common substring kernel for similarity between the labels of the graph.
\end{itemize}

Table~\ref{tbl:comparative:all} gives the comparative results of our context models against the plain BoF representation of the source code corpus. Both of the contextual representations give results that are superior to the plain vector representation of documents. In some cases, the result is improved by $66\%$ in PD. Interestingly, the semantically enriched bag of identifier-context pairs outperforms the dependency-based context model. This could be due to the fact that dependency graphs must be considered in terms of the whole progam, comprising of inter-module call dependencies.

\begin{table}
\begin{center}
 \caption{The comparison of plain bag of identifiers with semantically enriched CV and DG context models} 
         \begin{tabular}{ p{1.9cm} c c p{1.7cm}   p{1.7cm}   p{1.7cm}   p{1.7cm}  }
    \hline
    \multirow{2}{*}{\textbf{System}} & \multicolumn{2}{c}{\textbf{BoI}}  &  \multicolumn{2}{c}{\textbf{$SSK_1$}}  &  \multicolumn{2}{c}{\textbf{$SSK_2$}} \\\cline{2-7}
    & \textbf{PD}  &  \textbf{TED}  & \textbf{PD}  &  \textbf{TED}  & \textbf{PD}  &  \textbf{TED}  \\ \hline
 Apache Ant&$21218$&$2287$&$8738$ $(-58.82\%)$&$2049$ $(-10.41\%)$&\cellcolor[gray]{0.8}{$7287$ $(-65.66\%)$}&\cellcolor[gray]{0.8}{$2014$ $(-11.94\%)$}\\ \hline 
 Apache Hadoop&$4452$&$743$&$2369$ $(-46.80\%)$&\cellcolor[gray]{0.8}{$682$ $(-8.21\%)$}&\cellcolor[gray]{0.8}{$2282$ $(-48.74\%)$}&$715$ $(-3.77\%)$\\ \hline 
 Apache Log4j&$2263$&$487$&$1112$ $(-50.86\%)$&$432$ $(-11.29\%)$&\cellcolor[gray]{0.8}{$1083$ $(-52.13\%)$}&\cellcolor[gray]{0.8}{$425$ $(-12.73\%)$}\\ \hline 
 Eclipse JDT Core&$31327$&$2748$&\cellcolor[gray]{0.8}{$11057$ $(-64.71\%)$}&\cellcolor[gray]{0.8}{$2235$ $(-18.67\%)$}&$13753$ $(-56.10\%)$&$2492$ $(-9.32\%)$\\ \hline 
 JDOM&$878$&$273$&\cellcolor[gray]{0.8}{$640$ $(-27.11\%)$}&\cellcolor[gray]{0.8}{$247$ $(-9.52\%)$}&$664$ $(-24.41\%)$&$257$ $(-5.86\%)$\\ \hline 
 JEdit&$9148$&$1286$&\cellcolor[gray]{0.8}{$5305$ $(-42.01\%)$}&\cellcolor[gray]{0.8}{$1232$ $(-4.20\%)$}&$6601$ $(-27.84\%)$&$1233$ $(-4.12\%)$\\ \hline 
 JFreeChart&$12444$&$1464$&\cellcolor[gray]{0.8}{$5618$ $(-54.85\%)$}&\cellcolor[gray]{0.8}{$1352$ $(-7.65\%)$}&$7109$ $(-42.87\%)$&$1414$ $(-3.42\%)$\\ \hline 
 JHotDraw&$3002$&$665$&\cellcolor[gray]{0.8}{$1946$ $(-35.16\%)$}&\cellcolor[gray]{0.8}{$651$ $(-2.11\%)$}&$2454$ $(-18.26\%)$&$654$ $(-1.65\%)$\\ \hline 
 JUnit&$1098$&$300$&\cellcolor[gray]{0.8}{$697$ $(-36.53\%)$}&\cellcolor[gray]{0.8}{$288$ $(-4.00\%)$}&$984$ $(-10.39\%)$&$304$ $(+1.33\%)$\\ \hline 
 Weka&$18023$&$2304$&\cellcolor[gray]{0.8}{$7362$ $(-59.15\%)$}&\cellcolor[gray]{0.8}{$1992$ $(-13.54\%)$}&9262 $(-48.61\%)$&$2010$ $(-12.76\%)$\\ \hline 
	\end{tabular}
\label{tbl:comparative:all}
\end{center}
\end{table}                             

\subsection{JEdit Case Study}

To demonstrate the usefulness of semantically enriched context models for both software modularization and program comprehension, we perform semantic clustering on JEdit using the choices we established in the previous section. We show how by taking a knowledge-rich method, one may obtain more meaningful results. Several software engineering activities involve computing the similarity between source code identifiers (terms), such as code abbreviation expansion, code search, and topic identification for program comprehension.  Knowing terms relatedness allows for discovering prevalent themes that run through the source code, as well as finding relevant documents or terms based on a user's search query.   

In this section, we perform a textual analysis on the JEdit project using the contextual representations. The JEdit editor consists of $536$ classes spread through $29$ packages. The  vocabulary consists of $5935$ distinct identifier names. In the first application, we demonstrate how identifier similarity can be employed to derive topics that run throughout the source code, followed by performing a context analysis to show the types that are prevalent in each package in the source program.

\subsection{Topic Analysis}
Similar to document clustering, clusters of terms are those that frequently co-occur in documents. Term clusters (from now on, referred to as topics) can be potentially used for the identification and representation of topics prevalent throughout the source code. To compute the proximity between identifiers, we use the semantically enriched bag of identifier-context representation. The new perspective is to consider the collection of identifier-context pairs per document as identifier versus document-context pairs, and count the number of times that an identifier occurs in terms of the occurrence of a type in a document. We use the normalized euclidean distance to compute pairwise distances between the identifier names from the enriched bag of document-context pairs.  Alternatively, it is possible to compute term-term similarity from the dependency-based context model. In this approach, the dependency graphs of all modules are merged to construct a whole-program dependency graph, comprising of all the identifiers occurring throughout the source program combined with their data dependency relations. Diffusion kernel can be employed to compute the similarity between the nodes (identifiers) in the dependency graph.

The produced distance matrix from the contextual vector representation is used to produce topics and is compared against the topics produced using the plain Bag-of-Identifiers representation (BoI). The clustering algorithm used here to compute the term clusters is a `complete' hierarchical algorithm.  Although we have opted for an agglomerative algorithm to group together terms, both the choice of partitioning algorithm as well as the method used can be debated, but this is out of the scope of this paper. The produced dendrogram is then cut into an priori known number of topics, to find groups of similar terms. %

After the examination of the documentation and running the application, we have identified $7$ topics within the software system. We have also verified the identifier names and topics by scanning through the relevant parts of the source code. The following topics are in close alignment with those identified by Kuhn et al.~\cite{kuhn:2007:semantic}, in which they apply their approach on JEdit to identify topics that run through the system. Some of the well-encapsulated topics include:
\begin{itemize}
\item Core domain concepts such as management of file system and user interactions. 
\item User interface including UI components and their layout.
\item Text Area with functionalities for improving the programming experience including bracket matching, auto-indentation and commands for commenting out code.%
\item Plugins for extension and adding features to the editor.
\item Search and Replace using both literal and regular expressions.
\item Regular expressions to specify inexact search and replace. 

\item BeanShell scripting language for tasks such as writing macros.

\end{itemize}

To make a comparison between the topics discovered from a plain representation of BoI and those from the enriched contextual representation with semantic knowledge, we have identified $6$ identifier names that capture different conceptual information about each topic. The selected names for each topic comprise of concepts associated with each topic. Table~\ref{tbl:top:labels} lists the set of labels for each topic. All the labels are identifier names that appear in the source code corpus and as such do not always correspond to topical terms, such as a \textit{text-editor} or \textit{regular-expression}. It is the collection of names grouped together in a cluster that can be used to analyze and interpret the topic associated with that cluster. 

\begin{table}
\begin{center}
 \caption{The labels for each topic of JEdit } 
         \begin{tabular}{ l  p {10cm} }
    \hline
   \textbf{Topic}   &  \textbf{Labels}\\ \hline
    	BeanShell  & BSHINIT, isJavaBaseAssignable, Interpreter, invoke, resolveJavaMethod, isWrapperType \\ \hline
    	 Search \& Replace  & doBackwardSearch, doForwardSearch, searchField, hyperSearch, replace, replaceSelection \\ \hline
    	Regular Expressions   & escapeRule, startRegexp, endRegexp, pattern, terminateChar, matchType  \\ \hline
    	Text Area & caretLine, autoIndent, findMatchingBracket, getLineCount, caret, getSelectedText \\ \hline
     	Core  & settingsDirectory, queueAWTRunner, createVFSSession, invokeAction, buffer, handleMessage \\ \hline
     	Plugins  & activatePlugin, author, description, version, pluginSet, getPluginJAR \\ \hline
 		UI & processKeyEvent, toolbar, menubar, needFullRepaint, focusedComponent, addDockableWindow \\ \hline
	\end{tabular}
\label{tbl:top:labels}
\end{center}
\end{table}

\begin{figure*}[t]
\begin{center}
\includegraphics[width=12.5cm]{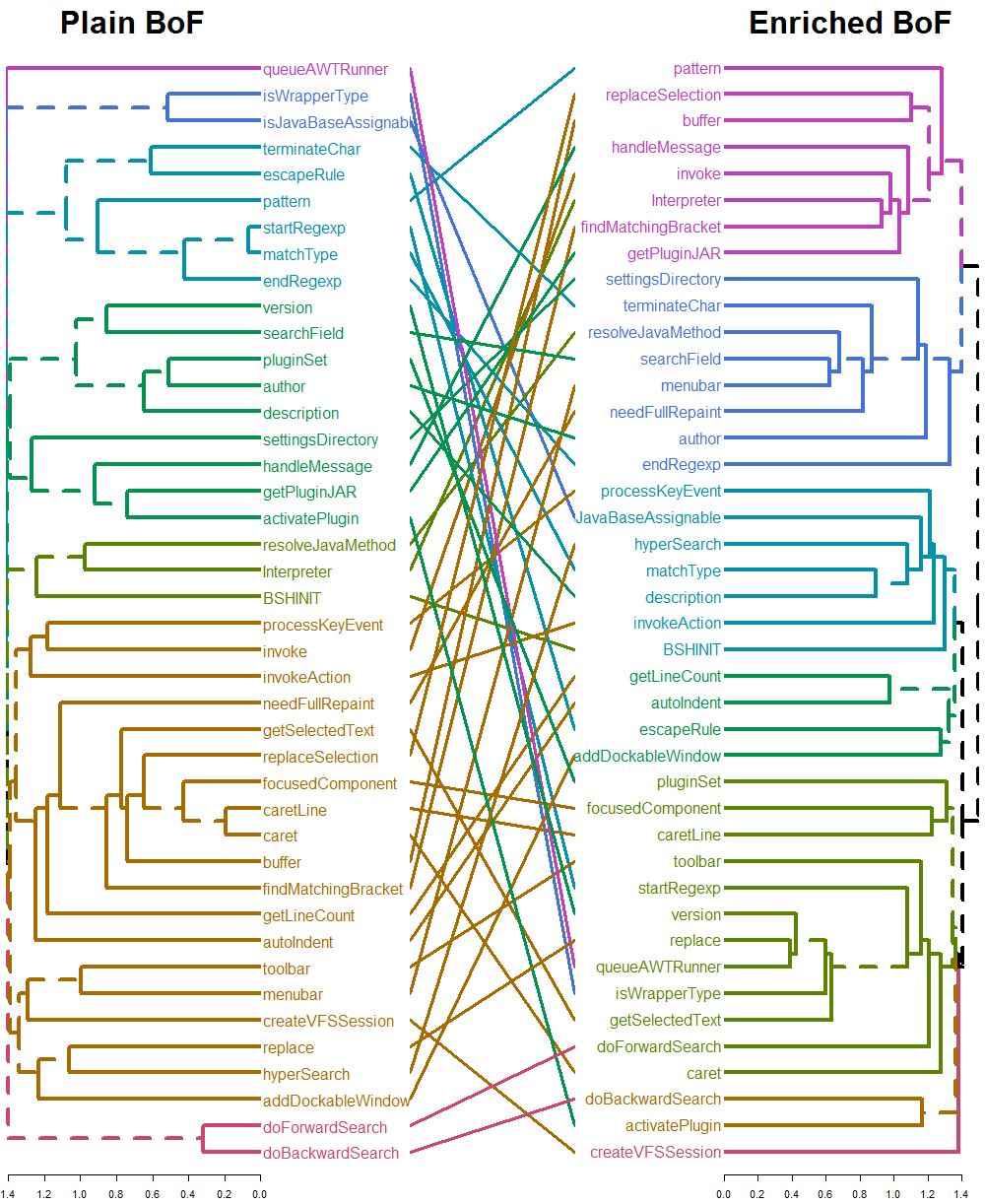}
\caption{The topics of JEdit and their labels}
\label{fig:topics}
\end{center}
 \end{figure*}

\autoref{fig:topics} depicts the side-by-side clusters (topics) of the identifier names for JEdit and its labels, both for the plain BoF and the enriched contextual vector representation. We first make a quantitative comparison against the established ground truth by evaluating the quality of produced clusters against the reference topic using the \emph{TED} and \emph{PD} scores. The topics discovered from the plain representation has $103$ unit TED and $237$ PD difference with the reference topic decomposition, whereas the enriched variant gives $91$ unit TED and $212$ PD difference. This demonstrates that by enriching the vector representation of document with semantic information, it is possible to obtain more meaningful results for topic identification in the source code.

Examining the tanglegram shown in \autoref{fig:topics} demonstrates that the produced topics for the enriched contextual variant tends to be more coherent, and align better with those of the reference topic decomposition. On the other hand, the topics discovered using the plain representation tends to be more heterogeneous. In the case of regular expressions, both decompositions seem to exhibit similar grouping.  

\subsection{Context Analysis}
In another application of contextual representation, we investigate the relations between the software system and the prevalent contexts throughout the source program. We will use the module-type representation of the source program to identify how the types are used in each package, and use that information to group together the types as well as the source modules. We have grouped together the package structure of JEdit system into $10$ packages (while merging packages deep in hierarchy and eliminating packages with fewer than 5 classes). We have used the normalized euclidean distance to compute the distances between the types as well as modules. Since each class is a referenceable type that can be used from within the program, the types consist of references to classes that are part of the source program.

\begin{figure*}[t]
\begin{center}
\includegraphics[width=11.2cm]{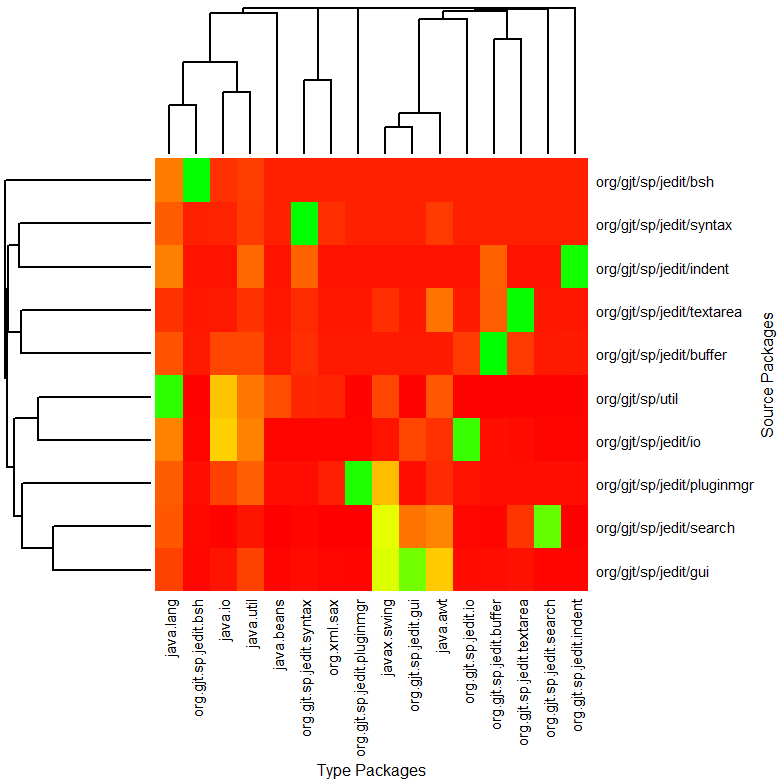}
\caption{The types used in JEdit}
\label{fig:types}
\end{center}
\end{figure*}

Figure~\ref{fig:types} illustrates the heatmap corresponding to the relations between the modules confined in the packages and the types used in the source program. The color red indicates a lack of relation, whereas light green indicates strong similarity. As shown, the utility types such as the ones imported from `java.util' are commonly used throughout the program. The gui package of jEdit uses types from libraries `javax.swing' and `java.awt'. Another observation is that each package mainly depends on the types internally encapsulated within that subdirectory, indicating that the packages are well-defined. This observation further supports our argument for using the package structure as the authoritative decomposition.

\subsection{Discussions}
Although enriching identifiers with semantic information helps to better understand a program, a more rigorous approach needs to be employed to fully exploit the underlying formal semantic framework. We need to distinguish between different zones in the source code, while treating identifier names as lexical objects with specific formal semantics, while comments can be treated as short texts in natural language.

The identifier names used for empirical evaluation are crude, however, a preprocessing step can be introduced for normalizing the names. As part of preprocessing, one can eliminate identifier names that are common words in a specific language, or terms that denote some common utility library (eg. bufferReader). Lawrie and Binkley~\cite{lawrie:2011:normalize} suggest an expansion algorithm to normalize the source code vocabulary. Another approach that can be accommodated in this process is that of Corazza et al.~\cite{corazza:2012:linsen}, called LINSEN. They propose to use an approximate string matching technique, and several general and domain specific dictionaries, to split identifiers and expand abbreviations.

Although a good naming convention and well-chosen identifier names help with understanding latent topics within source code, our approach is almost independent of the naming convention. By enriching the identifier names with formal semantic knowledge, it is possible to avoid problems with poor naming quality of identifiers. In case where names are cryptic abbreviations (as is the case with many legacy systems), this approach opens up new possibilities to understanding the topics and decode the names in the context of their co-clustered names.

\subsection{Threats to Validity}
The generalizability of our findings in this paper is limited by the restricted set of projects comprising of $10$ open source Java projects. We believe our techniques should be evaluated on more and larger systems to evaluate the effectiveness of our approach.

A major threat to the validity of this research is constructing the gold standard. In this paper, we have opted for the package structure of the software system to build the authoritative decomposition. This approach is in line with previous works~\cite{beck:2010:evolution, tzerpos:1999:mojo}, where they have also used the package structure for evaluation of their approaches. To ensure the quality of the ground truth, we have selected projects which are well-engineered or have gone through a re-structuring from the previous version. The results obtained by comparing the approaches with the authoritative decomposition is in alignment with our hypotheses, which further supports the quality of the ground truth. Furthermore, we have eliminated small packages with fewer than $4$ classes, and manually split larger packages with more than $40$ classes into smaller packages. Eliminating small and large packages ensure that the oracle decomposition itself doesn't exhibit an extreme distribution, and is uniformly grouped.

\section{Conclusion and Future Work}
\label{sec:conclusion}
In general, our results demonstrate the benefits of both introducing contexts for identifiers as well as enriching them with semantic knowledge. As demonstrated, in all cases, the semantically enriched context models give more authoritative decompositions when performing cluster analysis. Furthermore, as shown in the JEdit case study, the topics discovered are superior in quality. Our approach to evaluation of our technique is fully unsupervised, but it is possible to introduce supervision for filtering types, and selecting sensible identifier names to reduce noisy features when performing program comprehension analyses.

In the dependency-based context model, we restricted ourselves to flow-insensitive data dependencies. Making the data dependencies flow sensitive may further improve the results for this context model. Furthermore, in this representation, the context can be further refined by annotating each identifier with its type information. Our approach in enriching the data dependency graph of each module with semantic information about identifiers paves the way to perform semantically enriched fine-grained program analyses such as program slicing. Another thesaurus-based method for computing similarity is the gloss-based method. In general, members of the Public API are accompanied with documentation, which can be considered as a description of the member (i.e. gloss). We would like to integrate this source of knowledge to further improve semantic relatedness between source code identifiers.

\printbibliography

\appendix
\section{Conceptual Similarity}
\label{apx:sec:conceptual} 
\subsection{Knowledge Representation}
Before detailing our approach to enrich the identifiers in the source code, we first need to give some insights into approaches in representing semantic information about words in natural languages. One such approach is based on WordNet, which closely resembles our approach to knowledge representation. WordNet is a network of words with each playing a different role in the structure. In WordNet, words are grouped into blocks of what is known as a \emph{synset} (a synonymous set). A synset is a set of semantically synonymous words denoting the same concept. Each word may take a different sense\footnote{We will use the term `sense' to denote the meaning/purpose of the identifier}, depending on the context in which it is used, and hence, it may belong to more than one synset. For example, in the sentences `A tree is tall' and `A tree is an ADT', the concept \emph{plant} is intended by the usage of `tree' in the first sentence, whereas `tree' in the second sentence is a \emph{data structure}.

The synsets are interconnected with different relational links, such as \emph{hypernymy}, \emph{meronymy}, and \emph{synonymy}. Some of the types of relationships that can be found in a WordNet ontology are as follows:

\begin{itemize}
\item Synonymy: denotes an equivalence relation between two terms or concepts. Couch and sofa are two synonymous terms. 
\item Polysemy: A term that has different meanings depending on the context it is used. A tree has a different meaning in computer science and botany.
\item Hyper/Hyponymy: denotes a hierarchical relation between two terms. For instance, a dog (hyponym) is a type of an animal (hypernym).
\item Holo/Meronymy: A composite relationship between a whole and a part. For example, a city (meronym) is part of a country (holonym).
\item Instance-of relation: designates the relation between a general concept and an individual instance of that concept. For example, Java is an instance of a programming language.

\end{itemize}

Hyper/hyponymy and holo/meronymy are hierarchical relations, whereas the Instance-of relation is an associative relation. Synonymy and polysemy are symmetric relations.
 
\subsection{Semantic Network}
As stated before, the semantic relationships between terms and concepts form a hierarchical structure which can be represented in terms of a semantic network. A semantic network like WordNet is a scheme for representing semantic relations within a lexicon. The network is represented using a weighted directed graph with labelled nodes representing the words in the lexicon that appear in the corpus of the source code, and links denoting different types of relationships between them. The weights of each edge denotes the frequency of a relationship, extracted from the background information.

We will represent the semantic information about the software system in terms of a semantic network, where nodes are indexed by unique features (i.e. identifier names) in the source code, and edges represent the association strength between different nodes. A \emph{type} is defined by the set of its instances, forming a synonymous group. For example, `tree1' and `tree2' both being instances of type `Tree', are semantically synonymous. The \emph{instance-of} (ISA) relationship is used to link instances to types (concepts). The \emph{is-a-type-of} (ITO) denotes a hyponymy relationship between concepts. Each concept may contain properties or attributes that all its derivations will inherit. The fact that an animal has skin and color, implies that birds also have those properties.  The meronymy relationship from the parts to its whole is represented using a \emph{is-part-of} (IPO) link. The ISA and ITO relationship are transitive. For instance, a canary is a type of a bird, but bird is a kind of an animal, hence, a canary is an animal too.
 
To further clarify how the knowledge about the exposed members of a class would be represented in terms of a semantic network, let us consider the extended version of the example, given in \autoref{lst:employee:snippet}.
 
\begin{lstlisting}[caption=An example in Java, frame=tlrb, label=lst:javaexample, basicstyle=\fontsize{8}{9}\selectfont\ttfamily]{Name}

public abstract class Vehicle { 
    public int gear;
    public int speed;

    public Vehicle (int startSpeed, int startGear) { ... }

    public void setGear(int newValue) { ... }
        
    public void applyBrake(int decrement) { ... }
        
    public void setSpeed(int speed) { ... }
}

public class Car extends Vehicle{
    public Car(int startSpeed, int startGear) { ... }       
}

public class Employee{
    public String name;
    public Car car; 
        
    public Employee(String name, Car car) { ... }  
    
    public void raiseSalary(double byPercent, double bonus) { ... }  
}

\end{lstlisting}
 
\begin{figure}
\begin{center}
\includegraphics[width=9cm]{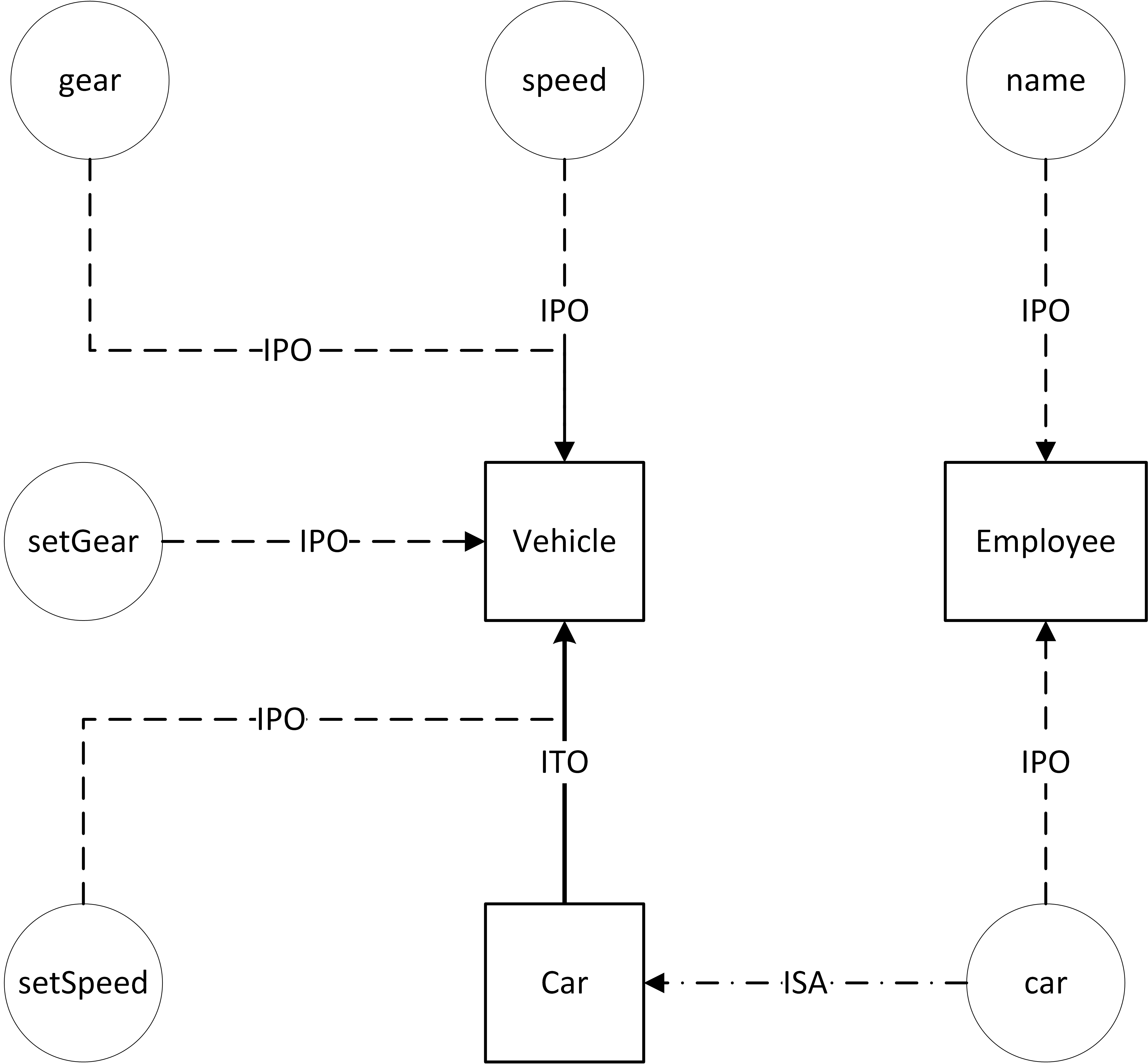}
\caption{The semantic network of the Java example program, composed of abstract nodes designating the identifier names and abstract links designating different types of relationship between the nodes.}
\label{fig:SN}
\end{center}
 \end{figure}

Figure~\ref{fig:SN} depicts the semantic network for the example program. There are two distinct kinds of nodes: terms denote identifier names, whereas concepts correspond to types.  For each concept, a synset exists comprising of the type itself and its instances.  For example, the identifier names, \emph{gear}, \emph{speed}, \emph{setGear}, \emph{applyBrake}, and \emph{setSpeed} form a synset group.  Here, the primitive type `int' and standard library type `String' are omitted. Notice that even though there are no direct links between the term \emph{car} and concept \emph{Vehicle}, traversal of the ITO link from \emph{Car} to \emph{Vehicle} induces a relationship between them.

Throughout the rest of this section, we restrict ourselves to object-oriented languages with a static type system (here, Java) and use the program's type information to compute an approximation of the semantics of the identifier. However, some of our techniques, as will be elaborated later can be used in the context of other programming paradigms.

\subsection{Semantic Relatedness Measurements}
Before we can give different semantic relatedness measurements, we briefly discuss some concepts to better understand various similarity measurements.

\subsubsection{Path Length}
Path length measurements are based on edge counting methods. A simple edge counting method is by Rada et al.~\cite{rada:1989:ranking}, which defines the conceptual distance between two nodes in the ITO network as the shortest path connecting the two nodes. Two major drawbacks of this approach is that 1) most semantic networks have a non-uniform density of edges, and 2) it doesn't take into account other semantic relationships like holo-meronymy. 

\begin{figure}[t]
\begin{center}
\includegraphics[width=8cm]{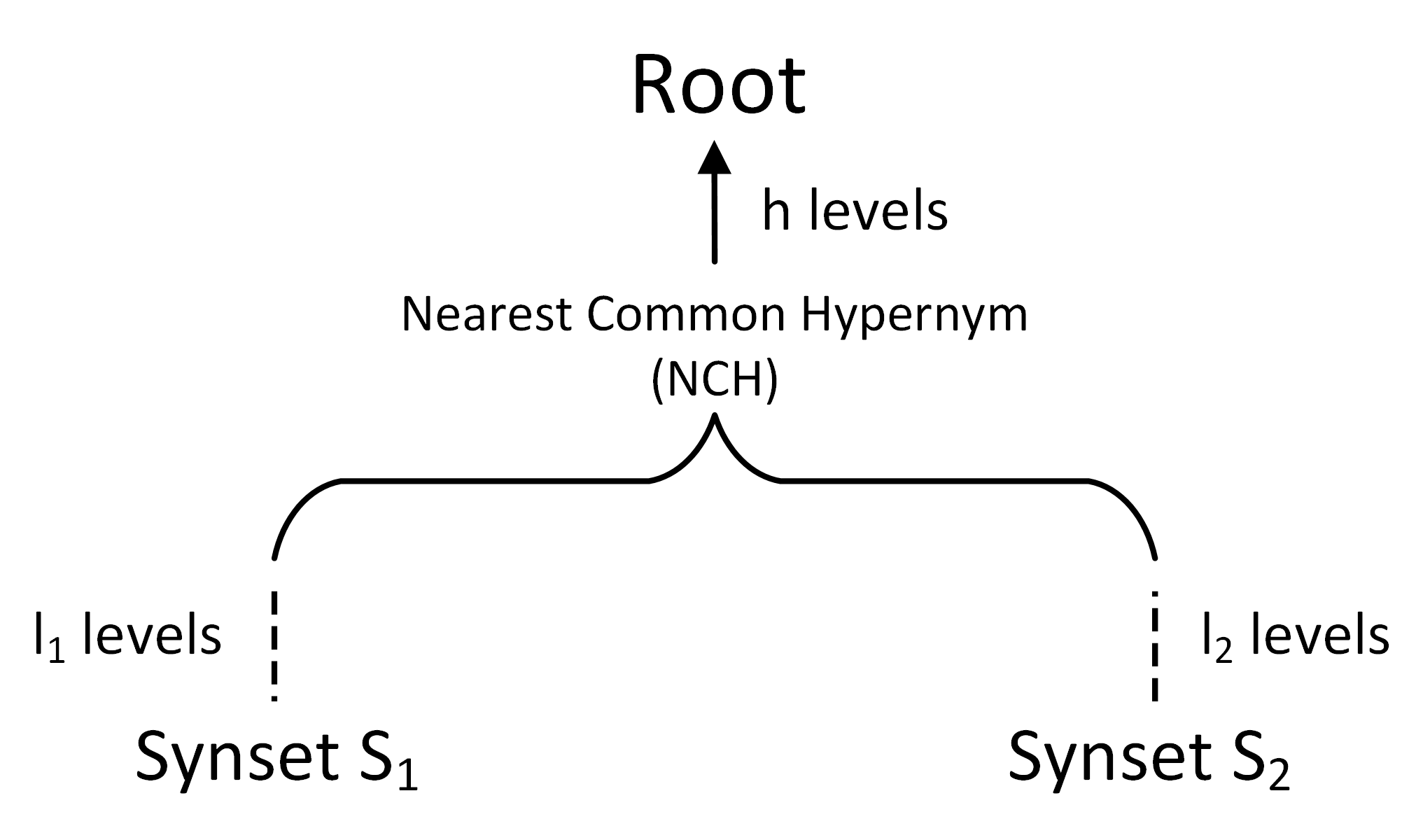}
\caption{Notations for inheritance structure}
\label{fig:measurements}
\end{center}
 \end{figure}

\autoref{fig:measurements} illustrates different notations used to describe various semantic similarity measurements. The depth of a type is its distance to the unique root node\footnote{If the structure is a directed graph, the minimal depth is considered.}. We assume a virtual top node (for instance, \textbf{java.lang.Object}) that dominates all nodes. The \textit{nearest common hypernym} (\textit{nch}) of two types refers to the type with the maximal depth that subsumes both types. The length of the shortest path from $T_1$ to $T_2$ is $d(T_1, T_2) = l_1 + l_2$.

\subsection{Semantic Similarity Measurements amongst Types}

Semantic similarity measures were initially defined based on shortest path length measure, and subsequently new measurements were proposed based on different design principles. In the following, we present various similarity scores for computing the semantic similarity between the types in the semantic network. 

\subsubsection{Inverted Path Length}
The Inverted Path Length (IPL) is a path length-based measurement to compute similarity, given in \autoref{eqn:IPL}, where $\alpha$ denotes the rate of decay. Despite its simplicity, the IPL does not take into account that types closer to the root of the hierarchy should have a lower weight than the ones in lower levels in the hierarchy.

\begin{equation}
sim_{IPL}(T_1, T_2) = \frac{1}{(1+d(T_1, T_2))^\alpha}
\label{eqn:IPL}
\end{equation}

\subsubsection{Wu \& Palmer}
Wu and Palmer~\cite{wu:1994:verbs} describe another path-length based measurement, given in \autoref{eqn:WUP} that scales the path-length with respect to the shortest path between the two types $T_1$ and $T_2$ to their closest common ancestor (NCH) and the depth of NCH to the unique root node.
\begin{equation}
sim_{WUP}(T_1, T_2)= \frac{2dep(nch(T_1, T_2))}{d(T_1,nch(T_1,T_2)) + d(T_1, nch(T_1, T_2) + 2dep(nch(T_1, T_2))}
\label{eqn:WUP}
\end{equation}

\subsubsection{Leacock \& Chodorow}
The Leacock and Chodorow measure~\cite{leacock:1998:wordnet} is another path-length measurement that computes the similarity score between two types $T_1$ and $T_2$ as the shortest path between them, divided by double the maximal depth of the two types in the taxonomy.

\begin{equation}
sim_{LC}(T_1, T_2) =  - log \frac{d(T_1, T_2)}{2 max\{depth(T_1),depth(T_2)\}}
\label{eqn:LC}
\end{equation}

\subsubsection{Conceptual Density}
An alternative to IPL is the Conceptual Density (CD)~\cite{agirre:1996:wsd} semantic similarity measurement which tries to scale the score by taking into account the topological structural of the semantic network. The CD measurement is still sensitive to the length of the shortest path that connects the types, but the types in a deeper part of the hierarchy are ranked closer. Furthermore, the similarity score is normalized to make it compatible with the density of the types: the types in a dense part of the hierarchy are ranked closer than the ones in a more sparse region~\cite{agirre:1996:wsd}.
\begin{equation}
sim_{CD}(T_{1}, T_{2})=\frac{\Sigma_{i=0}^{h}(\mu(\overline{nch(T_1,T_2)}))^i}{|\overline{nch(T_1,T_2)}|}
\label{eqn:CD}
\end{equation}
\newcommand{\floor}[1]{\lfloor #1 \rfloor}
where, for $g=nch(T_1,T_2)$, %
\begin{itemize}
\item $\mu(\overline{g})$ is the average number of children per node (i.e. the branching factor) in the sub-hierarchy of the nearest common hypernym of the types $T_1$ and $T_2$.
\item $h$ is the depth of the ideal, i.e. maximally dense, tree with enough leaves to cover the two types, $T_1$ and $T_2$, based on the average branching factor of $\mu(\overline{g})$.
This value is estimated by:
$$
h=
\begin{cases}
\floor{\log_{\mu(\overline{g})} 2} & \text{iff }\mu(\overline{g}) \neq 1  \\  2 &  \text{ otherwise}\   
\end{cases}
$$	
When $\mu(g)=1$, h ensures a tree with at least 2 nodes to cover $T_1$ and $T_2$ ($height = 2$).
\item $|\overline{g}|$ is the number of nodes in the sub-hierarchy $\overline{G}$. %
\end{itemize}

\subsection{Word Sense Disambiguation}
Each identifier name may have multiple senses (belong to different synsets). To disambiguate between different senses of an identifier, for each pair of identifiers, we choose their senses such that the similarity score between them is the highest. Since the network is a weighted graph, in case of ISA relationship, we normalize the score by the outdegree of the types in each sense. 
Here we define the similarity of two identifiers  $id_1$ and $id_2$ as: 
\begin{equation}
sim(id_1,id_2) = max_{1 \leq i \leq m, 1 \leq j \leq n}  sim(T_{ai} , T_{bj}) \times (\frac{w(id_1, T_{ai})}{2out(T_{ai})} +\frac{w(id_2, T_{bj})}{2out(T_{bj})})
\end{equation}
where, $T_{ai}$ and $T_{bj}$ belong to synonym set $S_{ai}$ and $S_{bj}$, respectively.
$S_{a1}, \ldots , S_{am}$ are the synsets of $id_{a}$, and $S_{b1}, \ldots , S_{bn}$ are the synsets of $id_{b}$.

\subsection{Diffusion of Similarity}

All the measurements between the types in the first approach depend on the ISA structure of the type hierarchy. We believe that the ISA structure of the type system does not fully capture all the relations between types. 
\begin{enumerate}
\item The problem with measurements based on shortest-path distance on the graph is that i) mere reachability between two nodes is not a good indicator of measurement between two nodes, and ii) it is extremely sensitive to the insertion and deletion of individual edges.

\item The properties/attributes contained within a type are neglected. However, these elements can be used to construct access paths to access properties in another type or invoke operations within that type, implicitly denoting a semantic relation from the source to the target. 

\item Common types (utility and third-party libraries) should have lower weight compared to application types. Treating the semantic network as a graph, less similarity is diffused through nodes with high outdegree.

\end{enumerate}

We employ a diffusion graph kernel, as outlined in \autoref{appendix:sec:graphkernels} to compute the global similarity between any two nodes in the semantic network. A diffusion process helps to build similarity between identifier names that are not directly related. When computing this kernel, we do not distinguish between different types of relationships between nodes in the semantic network. Hence, this approach can be employed in the context of programming languages with no type system, such as Cobol.

\section{Lexical Similarity}
\label{apx:sec:lexical}
Although identifiers have formal semantics, their names also carry meaningful information that needs to be exploited. For instance, identifiers  `carModel' and `carOwner'  are of different types, yet both involve holding information about the `Car' entity. Here, we have employed three techniques presented in \autoref{tbl:lexical:similarity:measurements} to induce similarity between identifier names. One technique is based on the length of the Longest Common Subsequence (LCS) of two identifier names normalized by the product of their respective lengths. The longest common subsequence of two strings is the maximal number of characters that is common to both of them. For instance, the LCS of the strings `carOwner' and `carModel' is `caroe'.
Another technique involves dividing the length of the Longest Common Substring (LCU) of two identifiers by the product of the length of each name. The LCU of the two identifier names `carOwner' and `carModel' is `car'.
The last technique used for conducting experiments is a string kernel using suffix arrays called Constant (Const) string kernel  which matches all common substrings between two strings and weights them equally.

\begin{table} 
\centering 
\caption{Measures of lexical similarity between identifier names}
\begin{tabular}{lc}

Longest Common Subsequence&\\
\multicolumn{2}{c}{$sim_{LCS}(id_1, id_2) =  \frac{(LCS(id_1, id_2))^2}{|id_1|.|id_2|}$}\\ \\
Longest Common Substring&\\
\multicolumn{2}{c}{$sim_{LCU}(id_{1}, id_{2})= \frac{(LCU(id_1, id_2))^2}{|id_1|.|id_2|}$}\\\\
Constant String&\\
\multicolumn{2}{c}{$sim_{Const}(id_{1}, id_{2})= \sum_{s \in id_{1}, s' \in id_{2} } num_{s}(id_{1}) num_{s}(id_{2})$}\\\\
\end{tabular}
\label{tbl:lexical:similarity:measurements}
\end{table}

\section{Designing a Semantic Kernel}
\label{apx:sec:semantickernel}
Kernels are used to incorporate a domain-specific notion of proximity in the input space. A kernel function $k(x_1, x_2)$ is used to compute the similarity between two objects $x_1$ and $x_2$ as a dot-product in a new vector space. An important property of kernel functions is that the exact mapping from the input space into this new vector space is not necessary (also known as the \emph{kernel trick}). Therefore, defining a good kernel means finding a similarity function that best captures the notion of similarity in that domain. In this paper, to compute similarity between source code documents (i.e. classes), we will use the linear kernel with \emph{cosine normalization modifier} (equivalent to cosine similarity). The reader is referred to the rich literature on kernel methods (eg.~\cite{scholkopf:2001:kernel, taylor:2004:kernel}) for further information.

We have adopted the approach from~\cite{wang:2008:wikipedia} to enrich the vector representation of documents. Embedding the semantic knowledge in the document vector requires transforming the vector representation of document $d$ by $\phi'(d) = \phi(d)P$, where P is a proximity matrix. Using the above transformation, the vector space kernel between two documents $d_1$ and $d_2$ becomes:

\begin{equation}
k(d_1, d_2) = \phi(d_1)PP^{\top}\phi(d_2)^{\top}
= \phi'(d_1) \phi'(d_2)^{\top}
\end{equation}

Hence, the new kernel is the inner product of the transformation of each document.

$P$ can be defined as $P=RS$, where $S$ defines the semantic similarities between features in the source code corpus, and $R$ is a diagonal matrix containing the term weighing, for which we will use the \textit{inverse term frequency} (idf) score to punish common feature names.  Embedding the proximity matrix into the vector space model (VSM) corresponds to representing a document in a less sparse vector, where there is a non-zero entry for all the terms that are semantically similar~\cite{wang:2008:wikipedia}. If there is no semantic relation between any two terms, no similarity can be induced from their co-occurrence.

\section{Evalutation of Enrichment Processes}
\label{apx:sec:evaluationenrichment}
Table~\ref{tbl:comprative:lexical} gives the results for different lexical similarity metrics. The string kernel computed for each metric comprises of the similarity of identifier names extracted from the body of methods in each document with the tf-idf weighing schema. As shown, the normalized longest common substring (LCU) yields the best result. 

\begin{table*}
\begin{center}
 \caption{The comparison of various lexical similarity measurements} 
         \begin{tabular}{  l  c  c  c  c  c  c }
    \hline
    \multirow{2}{*}{\textbf{System}} & \multicolumn{2}{c}{\textbf{LCS}}  &  \multicolumn{2}{c}{\textbf{LCU}}  &  \multicolumn{2}{c}{\textbf{Const}} \\\cline{2-7}
    & \textbf{PD}  &  \textbf{TED}   & \textbf{PD}  &  \textbf{TED}   & \textbf{PD}  &  \textbf{TED}  \\ \hline
  Apache Ant&$10437$&\cellcolor[gray]{0.8}{$2030$}&\cellcolor[gray]{0.8}{$9594$}&$2040$&$11626$&$2044$\\ \hline 
 Apache Hadoop&$2584$&$717$&\cellcolor[gray]{0.8}{$2381$}&\cellcolor[gray]{0.8}{$715$}&$2858$&$719$\\ \hline 
 Apache Log4j&\cellcolor[gray]{0.8}{$1551$}&\cellcolor[gray]{0.8}{$483$}&$1618$&$487$&$1740$&$487$\\ \hline 
 Eclipse JDT Core&$17059$&$2547$&\cellcolor[gray]{0.8}{$16410$}&\cellcolor[gray]{0.8}{$2536$}&$17236$&$2585$\\ \hline 
 JDOM&$710$&$250$&\cellcolor[gray]{0.8}{$691$}&\cellcolor[gray]{0.8}{$248$}&$859$&$256$\\ \hline 
 JEdit&$5923$&$1224$&\cellcolor[gray]{0.8}{$4946$}&$1236$&$6610$&\cellcolor[gray]{0.8}{$1218$}\\ \hline 
 JFreeChart&$6397$&$1438$&\cellcolor[gray]{0.8}{$5608$}&\cellcolor[gray]{0.8}{$1436$}&$7344$&\cellcolor[gray]{0.8}{$1436$}\\ \hline 
 JHotDraw&$2249$&\cellcolor[gray]{0.8}{$643$}&\cellcolor[gray]{0.8}{$2174$}&$647$&$2464$&$647$\\ \hline 
 JUnit&$721$&$289$&\cellcolor[gray]{0.8}{$672$}&$289$&$736$&\cellcolor[gray]{0.8}{$285$}\\ \hline 
 Weka&$11685$&$1952$&\cellcolor[gray]{0.8}{$9173$}&\cellcolor[gray]{0.8}{$1792$}&$9596$&$1867$\\ \hline
	\end{tabular}
\label{tbl:comprative:lexical}
\end{center}
\end{table*}

Table~\ref{tbl:comparative:semantic:similarities} gives the results for various semantic relatedness measurements. In general, the conceptual density (CD) measurement performs well in comparison with other methods. The CD measurement computes the proximitiy of the senses of the words based on the hypernymy relationship, as the information expressed by the maximally dense subhierarchy that includes both senses.
 
\begin{table*}
\begin{center}
 \caption{The comparison of various semantic relatedness measurements} 
  \begin{tabular}{  l  c  c  c  c  c  c  c  c  c  c }
    \hline
    \multirow{2}{*}{\textbf{System}} & \multicolumn{2}{c}{\textbf{IPL}}  &  \multicolumn{2}{c}{\textbf{WUP}}  &  \multicolumn{2}{c}{\textbf{LC}} &  \multicolumn{2}{c}{\textbf{CD}} \\\cline{2-9}
    & \textbf{PD}  &  \textbf{TED} &  \textbf{PD}  &  \textbf{TED}   &  \textbf{PD}  &  \textbf{TED}   &\textbf{PD}  &  \textbf{TED}   \\ \hline
  Apache Ant&$18509$&$2038$&$13454$&$2166$&$11965$&$2026$&\cellcolor[gray]{0.8}{$11468$}&\cellcolor[gray]{0.8}{$1965$}\\ \hline 
 Apache Hadoop&$3918$&$713$&$3344$&$721$&\cellcolor[gray]{0.8}{$3051$}&\cellcolor[gray]{0.8}{$709$}&$3146$&$723$\\ \hline 
Apache Log4j&$2277$&$485$&$2009$&$489$&\cellcolor[gray]{0.8}{$1895$}&\cellcolor[gray]{0.8}{$481$}&$2013$&$487$\\ \hline 
 Eclipse JDT Core&$19274$&$2643$&$18154$&$2612$&$18039$&$2602$&\cellcolor[gray]{0.8}{$17328$}&\cellcolor[gray]{0.8}{$2552$}\\ \hline 
 JDOM&$856$&$254$&$795$&$262$&$798$&$256$&\cellcolor[gray]{0.8}{$787$}&\cellcolor[gray]{0.8}{$248$}\\ \hline 
 JEdit&$8892$&$1244$&$6836$&\cellcolor[gray]{0.8}{$1220$}&\cellcolor[gray]{0.8}{$6649$}&$1228$&$6873$&$1234$\\ \hline 
 JFreeChart&$9624$&$1454$&$6539$&$1450$&\cellcolor[gray]{0.8}{$6146$}&\cellcolor[gray]{0.8}{$1430$}&$7125$&$1452$\\ \hline 
 JHotDraw&$2814$&$637$&$2139$&$645$&$2483$&$641$&\cellcolor[gray]{0.8}{$2129$}&\cellcolor[gray]{0.8}{$627$}\\ \hline 
 JUnit&$7968$&$289$&$808$&$289$&$865$&$285$&\cellcolor[gray]{0.8}{$706$}&\cellcolor[gray]{0.8}{$281$}\\ \hline 
 Weka&$14716$&$2451$&$13488$&$2254$&$12152$&$2098$&\cellcolor[gray]{0.8}{$11436$}&\cellcolor[gray]{0.8}{$2035$}\\ \hline 
	\end{tabular}
\label{tbl:comparative:semantic:similarities}
\end{center}
\end{table*}     

Based on the previous results, we have configured our enrichment approaches for both the string kernel and the choice of semantic relatedness measure. We compare the performance of the following three methods:                                                                                                                                                                            
\begin{itemize}
\item BoI (baseline): A bag of identifier names (BoI) extracted from the body of methods with the idf weighing schema.
\item $SSN_1$: The bag of features enriched with the semantic matrix from element-wise product of CD metric for semantic relatedness and normalized common substring kernel with the idf weighing schema.
\item $SSN_2$: The bag of features enriched with the semantic matrix from element-wise product of the diffusion kernel on the semantic network and normalized common substring kernel with the idf weighing schema.
\end{itemize}

Table~\ref{tbl:comparative:semantic:all} gives the comparative results of our enrichment approaches against the plain BoF representation of the source code corpus. Both of the enrichment approaches give results that are superior to the plain vector representation of documents. In some cases, the result is improved by $65\%$. As anticipated, the diffusion kernel-based semantic matrix outperforms the NLP-based semantic relatedness measures in $7$ out of $10$ cases for PD metric and $6$ cases for TED score.

\begin{table}
\begin{center}
 \caption{The comparison of identifier similarity measurements with plain bag of identifiers, and the enriched variants with CD and diffusion kernel} 
         \begin{tabular}{  m{1.9cm}   c  c m{1.7cm}  m{1.7cm}  m{1.7cm}  m{1.7cm}| }
    \hline
    \multirow{2}{*}{\textbf{System}} & \multicolumn{2}{c}{\textbf{BoI}}  &  \multicolumn{2}{c}{\textbf{$SSN_1$}}  &  \multicolumn{2}{c}{\textbf{$SSN_2$}} \\\cline{2-7}
    & \textbf{PD}  &  \textbf{TED}  &  \textbf{PD}  &  \textbf{TED}  & \textbf{PD}  &  \textbf{TED}  \\ \hline
Apache Ant&$21218$&$2042$&$8235$ $(-61.19\%)$&$1996$ $(-2.25\%)$&\cellcolor[gray]{0.8}{$7457$ $(-64.86\%)$}&\cellcolor[gray]{0.8}{$1961$ $(-3.97\%)$}\\ \hline 
 Apache Hadoop&$4452$&$743$&\cellcolor[gray]{0.8}{$2264$ $(-49.14\%)$}&\cellcolor[gray]{0.8}{$705$ $(-5.11\%)$}&$2466$ $(-44.61\%)$&$711$ $(-4.31\%)$\\ \hline 
 Apache Log4j&$2263$&$487$&$1534$ $(-32.19\%)$&$479$ $(-1.64\%)$&\cellcolor[gray]{0.8}{$1365$ $(-39.69\%)$}&\cellcolor[gray]{0.8}{$457$ $(-6.16\%)$}\\ \hline 
 Eclipse JDT Core&$31327$&$2748$&$16522$ $(-47.26\%)$&$2561$ $(-6.80\%)$&\cellcolor[gray]{0.8}{$14344$ $(-54.21\%)$}&\cellcolor[gray]{0.8}{$2374$ $(-13.61\%)$}\\ \hline 
 JDOM&$878$&$273$&\cellcolor[gray]{0.8}{$600$ $(-31.65\%)$}&\cellcolor[gray]{0.8}{$238$ $(-12.82\%)$}&$765$ $(-12.87\%)$&$265$ $(-2.93\%)$\\ \hline 
 JEdit&$9148$&$1286$&$4769$ $(-47.87\%)$&$1220$ $(-5.13\%)$&\cellcolor[gray]{0.8}{$4461$ $(-51.24\%)$}&\cellcolor[gray]{0.8}{$1194$ $(-7.15\%)$}\\ \hline 
 JFreeChart&$12444$&$1446$&\cellcolor[gray]{0.8}{$5474$ $(-56.01\%)$}&\cellcolor[gray]{0.8}{$1372$ $(-5.12\%)$}&$5652$ $(-54.58\%)$&$1380$ $(-4.56\%)$\\ \hline 
 JHotDraw&$3002$&$665$&$2028$ $(-32.42\%)$&\cellcolor[gray]{0.8}{$649$ $(-2.41\%)$}&\cellcolor[gray]{0.8}{$1958$ $(-34.77\%)$}&$653$ $(-1.80\%)$\\ \hline 
 JUnit&$1098$&$300$&$667$ $(-39.25\%)$&$287$ $(-4.33\%)$&\cellcolor[gray]{0.8}{$563$ $(-48.72\%)$}&\cellcolor[gray]{0.8}{$252$ $(-16.00\%)$}\\ \hline 
 Weka&$18023$&$2301$&$9773$ $(-45.77\%)$&$2114$ $(-8.13\%)$&\cellcolor[gray]{0.8}{$8471$ $(-53.00\%)$}&\cellcolor[gray]{0.8}{$2038$ $(-11.43\%)$}\\ \hline
	\end{tabular}
\label{tbl:comparative:semantic:all}
\end{center}
\end{table}

\section{Graph Kernels}
\label{appendix:sec:graphkernels}
Here, we outline two different sets of graph kernels, those used to compute the pairwise similarity between vertices in a graph, and the ones used to compute similarity between two graphs. For theoretical background and evaluation of different kernels on graphs, please refer to~\cite{fouss:2012:kernels}. 

\subsection{Diffusion Kernels}
A class of graph-based kernel functions are the diffusion kernels, originally introduced by Kondor and Lafferty~\cite{kondor:2002:diffusion}. The kernel function used in this paper is the exponential diffusion kernel~\cite{kondor:2002:diffusion}, defined as follows:
\begin{equation}
K_{EXP} = exp(\alpha A) = \sum^{\infty}_{k=0} \frac{\alpha^{k} A^{k}}{k!}
\end{equation}
where $\alpha$ denotes the sinking factor of each node and $A$ is the adjacency matrix of an undirected graph. This kernel represents an average of path counts between nodes, adjusted by the inverse factorial of path length.

\subsection{Random Walk Graph Kernels}
Random walk graph kernels are used to compute similarity between a pair of graphs. The generalized random walk graph kernel is based on the idea of random walks: given a pair of graphs, perform random walks on both, and count the number of matching walks.

We will use the Kronecker product of graphs $G_1$ and $G_2$. The product graph $G_\times = (V_\times, E_\times)$, is defined via
\begin{equation}
V_\times(G1 \times G2) = \{(v_1, w_1) \in V1 \times V2 \}
\end{equation}

For each pair of vertices $x = (v_1, w_1)$ and $y = (v_2, w_2)$ in $G_x$, the weighted adjacency matrix $A$ for the product of graphs is defined as below   
\begin{equation}
A_\times(x, y) = \sigma_{label}(v_1, w_1) \times A_1(v_1, v_2) \times A_2(w_1, w_2) \times \sigma_{label}(v_2, w_2)   
\end{equation}
where $\sigma_{label}$ is a function that computes similarity between the labels of two vertices.\\

The random walk kernel function is defined as follows:
\begin{equation}
k_\times(G_1, G_2) = \Sigma^{V_\times}_{i,j=1}[\Sigma^{\infty}_{n=0} \lambda^n A^n_{\times}]_{ij} = \mathbf{e}^{\top} ( \mathbb{I} - \lambda A_{\times} )^{-1} \mathbf{e}
\end{equation}

\end{document}